\documentclass[prb,a4paper,showpacs,superscriptaddress,english]{revtex4}
\usepackage{amsmath} 
\usepackage{amsfonts} 
\usepackage{amssymb} 
\usepackage{bm} 
\usepackage{graphicx}
\usepackage[latin1]{inputenc}
\usepackage{hyperref}
\usepackage{epstopdf} 
\begin{document}
\title{Fermi liquid theory applied to vibrating wire measurements in $^3$He-$^4$He mixtures}
\author{T. H. Virtanen}
\affiliation{Department of Physics,
P.O.Box 3000, FI-90014 University of Oulu, Finland}
\author{E. V. Thuneberg}
\affiliation{Department of Physics,
P.O.Box 3000, FI-90014 University of Oulu, Finland}
\date{\today}

\begin{abstract}
We use Fermi liquid theory to study the mechanical impedance of $^3$He-$^4$He mixtures at low temperatures. The theory is applied to the case of vibrating wires, immersed in the liquid. We present numerical results based on a direct solution of Landau-Boltzmann equation for the $^3$He quasiparticle distribution for the full scale of the quasiparticle mean-free-path $\ell$. The two-fluid nature of mixtures is taken into account in the theory, and the effect of Fermi liquid interactions and boundary conditions are studied in detail. The results are in fair quantitative agreement with experimental data. In particular, we can reproduce the anomalous decrease in inertia, observed in vibrating wire experiments reaching the ballistic limit. The essential effect of the experimental container and second-sound resonances is demonstrated. 
\end{abstract}

\maketitle

\section{Introduction}

Vibrating wire resonators have extensively been used to study the properties of quantum fluids. Different regimes can be observed in Fermi systems at low temperatures. At relatively high temperatures the fluid obeys hydrodynamic description and the vibrating wire can be used to extract the density and the coefficient of viscosity of the liquid. As the temperature is reduced below the Fermi temperature, the collisions between the quasiparticles, the elementary excitations of the liquid, get rare as the final states of a collision are limited by the Pauli principle. This leads to increasing mean-free-path of the quasiparticles, and to increasing viscosity. With increasing viscosity the vibrating wire becomes more sensitive to its surroundings since more liquid is dragged to move with the wire, and this becomes limited by the experimental container. The increasing mean-free-path also leads to deviations from the hydrodynamic behavior, which are first visible as a slip on the walls: the velocity of the fluid does not vanish at the wall but extrapolates to zero a slip-distance behind the wall (in the rest frame of the wall). 
Lowering the temperature further, the mean-free-path becomes comparable to the radius of the wire. In this regime the dynamics of the quasiparticle gas needs to be described by the Fermi liquid theory, which is similar as the theory of rarified gases but uses Fermi distribution and has remaining interactions between quasiparticles. In the limit of lowest temperatures one enters the ballistic limit where the mean-free-path exceeds the size of the experimental container. 

In this paper we provide theoretical calculations for a vibrating wire in the full range of mean-free-paths in a {\em normal} Fermi system. The theory is applicable to pure liquid $^3$He, but the range of applicability is limited by the superfluid transition. Therefore the main application is in $^3$He-$^4$He mixtures, where the fermions remain in the normal state in full range that has been measured. Several vibrating wire experiments have been made in $^3$He-$^4$He mixtures. We  compare our calculations to the measurements by Martikainen \emph{et al.}\cite{mart,Pentti09}, which extend deepest into  the ballistic regime.
In order to incorporate the $^4$He-component, we use Landau's Fermi-liquid theory \cite{Landau57}
generalized to the simultaneous presence of condensed bosons, as formulated by Khalatnikov\cite{khal}. A formulation of this theory adapted to the present work is given in Ref.\ \onlinecite{VTres}. A short account of the present calculations is given in Ref.\ \onlinecite{VTletter}. This letter highlights the Landau force, the macroscopic force caused by the Fermi liquid interactions. Here we calculate the Landau force among other effects and compare to experiments. We find strong effect of the experimental container on the resonance frequency of the vibrating wire. We also study sound resonances that affect the measurements at high frequencies or in large containers.   

Our calculation is the first two-dimensional solution of normal-state Fermi-liquid equations for general mean-free-path. Even the cylindrically symmetric case we study, although effectively one-dimesional, is far more sophisticated than has been done before. Previous work has considered uniform bulk liquid \cite{Landau57,AbrikosovKhalatnikov,LLs2,BaymPethick} or one-dimensional geometries such as an oscillating planar wall in semi-infinte liquid \cite{Bekarevich61,Flowers78,Richardson78} or Poiseulle flow between parallel planes\cite{HojgaardJensen80,EinzelParpia97}. For more general geometries calculations have been done only in limiting cases of small mean-free-path\cite{carl,bowley,Perisanu06} and in the ballistic limit \cite{gue2,bowley,VT,VT09}. 

We start in Sec.\ \ref{sec-theory} by stating more precisely the problem we study. We briefly present the equations we need, which are derived in detail in Ref.\ \onlinecite{VTres}. The symmetries of the problem are studied in Sec.\ \ref{sec-sym}. In Secs.\ \ref{sec-hd} and \ref{sec-bal} we study the special cases of vanishing mean-free-path in the hydrodynamic limit, and infinite mean-free-path in the ballistic limit. The numerical method is introduced in Sec.\ \ref{sec-num} and the parameter values used are discussed in Sec.\ \ref{sec-pv}. The results of the numerical calculations are presented and discussed in Sec.\ \ref{sec-res}.

\section{Formulation of the problem}\label{sec-theory}

\subsection{Main assumptions}\label{subsec-ma}

We study the force that a vibrating wire exerts on a Fermi liquid surrounding the wire.
The wire is modeled as an infinite  cylinder. We assume that both the wire and the liquid are in a stationary container that also has the shape of an infinite cylinder. These assumptions are not essential in principle, but they simplify the numerical calculations by allowing translation symmetry in the direction of the cylinder axis.  The cross section of the wire is taken as a circle of radius $a$.  
The oscillation direction of the wire (chosen as the $x$ axis) is perpendicular to wire axis (chosen as the $z$ axis).

For the container we consider two different shapes. By {\em cylindrical container} we mean a circular  cylindrical container of radius $b$ that is concentric with the wire. By {\em slab container} we mean that the wire is in the middle between two parallel plane walls spaced by $2h$, and the oscillation direction of the wire is normal to the walls. Because of the latter case, circular symmetry is not valid in general.

The force exerted on the liquid per unit length of the wire is denoted by $\bm F$.
We are interested in the linear response of the fluid. For that we express the force in terms of the mechanical impedance of the fluid, $Z$, as
\begin{equation}
{\bm F}=Z{\bm u},
\label{zdef}\end{equation}
where $\bm u$ is the velocity of the wire.
In general $Z$ is a tensor, but we concentrate on symmetric cases, where $Z$ is a scalar. We assume harmonic time dependence $\exp(-i\omega t)$ with angular  frequency $\omega$. Thus $Z$ is complex valued, $Z=Z'+iZ''$. The impedance is directly related to the resonance frequency $f_0$ and the line width $\Delta f$ of the resonator by
\begin{eqnarray}\label{Z-freq}
Z'&=&2\pi^2a^2\rho_w\Delta f,\nonumber\\Z''&=&4\pi^2a^2\rho_w(f_0-f_{\rm vac}).
\end{eqnarray}
Here $\rho_w$ is the density of the wire material and $f_{\rm vac}$ the resonance frequency of the wire in  vacuum. The restriction to linear response implies that the wire velocity is small compared to the Fermi velocity, $u\ll v_{\rm F}$, or any sound velocity in the fluid.
Because of limitation to linear response, we systematically write all equations below in their linearized form.

The plan is to calculate $Z$ in the full range  of the quasiparticle mean-free-path $\ell$, from the hydrodynamic regime $\ell\ll a$ to the ballistic limit  $\ell\gg a$. The liquid is assumed to be mixture of bosons and fermions, but the corresponding number densities $n_4$ and $n_3$ are arbitrary. Thus the theory applies to pure Fermi liquid as well ($n_4=0$).

Besides linearity, another simplification is that we neglect the coupling between the normal and superfluid components. This can be justified in two different limits: at small frequencies and at small concentrations. In the former case we aim to calculate $Z$ to accuracy that is correct to first order in $a\omega/v_{\rm F}$. The coupling contributes to $Z$ in second order in $\omega$. Assuming that the relevant dimensionless parameter is $a\omega/v_{\rm F}$, the coupling indeed is a small correction in the experimental case we study, where $a\omega/v_{\rm F}\sim 0.02$ (Table \ref{tab4}). In a finite container there appears a second length scale, $h= 8a$. Since $h\omega/v_{\rm F}$ is not very small, the small-frequency approximation is not necessarily accurate. For the superfluid component the low frequency approximation is well satisfied, $h\omega/c\ll 1$, due to the large sound velocity $c$ of liquid $^4$He. 
We note that one more frequency dependent parameter,  $a^2\omega/v_{\rm F}\ell$, occurs in the theory, but it can have any magnitude (see Sec.\ \ref{sec-hd}). 

Alternatively, the normal-superfluid coupling can also be neglected in the limit of small $^3$He concentration. This is because the response of the superfluid  to the motion of $^3$He is proportional to the ratio of their number densities, $n_3/n_4$
[\onlinecite{VTres}].   Thus, irrespective of frequency, the coupling can be neglected for small $n_3/n_4$.  

\subsection{Bulk Fermi-liquid theory}\label{sec-lin-eqs}
The Fermi liquid theory was formulated by Landau to describe the low energy states of interacting Fermi liquids\cite{Landau57}. It was generalized by Khalatnikov to simultaneous presence of condensed bosons\cite{khal}. A new presentation of the Fermi-Bose-liquid theory is given in Ref. \onlinecite{VTres}. Here we briefly review the equations of Ref. \onlinecite{VTres} that are necessary for the numerical calculation. 

The central quantity in the theory is the quasiparticle distribution function $\psi_{\hat{\bm p}}(\bm r,t)$. It depends on the momentum $\bm p$ only through its direction $\hat{\bm p}=\bm p/p$. In addition it depends on the location $\bm r$ and on time $t$. The distribution $\psi_{\hat{\bm p}}$ is obtained from the more common  quasiparticle distribution $n_{\bm p}$ by integration over the magnitude of the momentum and making a transformation that partly decouples the normal and superfluid components\cite{VTres}. The state of the bosons is described by the deviation of the chemical potential $\delta\mu_4(\bm r,t)=\mu_4^{ }(\bm r,t)-\mu_4^{(0)}$ from its equilibrium value  $\mu_4^{(0)}$ and by the superfluid velocity $\bm v_s(\bm r,t)$.

One more important quantity is the quasiparticle energy shift on the Fermi surface, $\delta\epsilon_{\hat{\bm p}}(\bm r,t)$. It depends on  $\psi_{\hat{\bm p}}$ and on $\delta\mu_4$ as
\begin{eqnarray}
\delta\epsilon_{\hat{\bm p}}=\frac{K}{1+ F_0}\delta\mu_{\rm 4} +\sum_{l=0}^\infty \frac{ F_l}{1+\frac{1}{2l+1} F_l} \langle P_l(\hat{\bm p}\cdot\hat{\bm p}')\psi_{\hat{\bm p}'}\rangle_{\hat{\bm p}'}. 
\end{eqnarray}
Here $K$ is a parameter describing coupling to the superfluid component,
$F_l$ with $l=0$, 1, 2, etc.\ are the Landau parameters describing interactions between quasiparticles,
$P_l$ are Legendre polynomials, and  $\langle \ldots\rangle_{\hat{\bm p}}$ denotes average over the unit sphere of $\hat{\bm p}$. It follows from Eqs.\ (\ref{e.lbrel2}) and (\ref{e.he4vefs}) below
that the coupling term contributes to $Z$ proportional to   $\omega^2$, and we neglect it. Since there are no experimental determinations  of $F_l$ for $l\geq2$, we neglect the corresponding terms. Therefore $\delta\epsilon_{\hat{\bm p}}$ simplifies to
\begin{eqnarray}\label{e.deltae2}
\delta\epsilon_{\hat{\bm p}}=\frac{F_0}{1+F_0}\langle \psi_{\hat{\bm p}'} \rangle_{\hat{\bm p}'}+\frac{F_1}{1+F_1/3}\hat{\bm p}\cdot\langle \hat{\bm p}'\psi_{\hat{\bm p}'} \rangle_{\hat{\bm p}'}.
\end{eqnarray}
The kinetic equation in the relaxation-time approximation takes the form
\begin{eqnarray}\label{e.lbrel2}
\frac{\partial}{\partial t}(\psi_{\hat{\bm p}}-\delta\epsilon_{\hat{\bm p}})+v_{\rm F}^{ }\hat{\bm p}\cdot\bm\nabla\psi_{\hat{\bm p}}=
-\frac{1}{\tau}(\psi_{\hat{\bm p}}-\psi^{\rm le}_{\hat{\bm p}}).
\end{eqnarray}
Here $v_{\rm F}$ is the Fermi velocity that is related to the Fermi momentum $p_{\rm F}$ and to the effective mass $m^*$ by $v_{\rm F}=p_{\rm F}/m^*$. In the collision term $\tau=\ell/v_{\rm F}$ is the relaxation time and the local-equilibrium distribution 
\begin{eqnarray}\label{e.ledfd}
\psi^{\rm le}_{\hat{\bm p}}=\langle\psi_{\hat{\bm p}'}\rangle_{\hat{\bm p}'}+3\hat{\bm p}\cdot\langle\hat{\bm p}'\psi_{\hat{\bm p}'}\rangle_{\hat{\bm p}'}.
\end{eqnarray}
Assuming time dependence $\exp(-i\omega t)$ and parameterizing the quasiparticle trajectories by ${\bm r}={\bm r}_0+s\hat{\bm p}$, the transport equation can be integrated in the form
\begin{equation}\label{e.lbrel2x}
\psi_{\hat{\bm p}}({\bm r}_0)=\psi_{\hat{\bm p}}({\bm r}_0+s_0\hat{\bm p})e^{ks_0} +\int_{s_0}^0ds\left[\frac{1}{\ell}\psi^{\rm le}_{\hat{\bm p}}({\bm r}_0+s\hat{\bm p})-i\frac{\omega}{v_F} \delta\epsilon_{\hat{\bm p}}({\bm r}_0+s\hat{\bm p})\right]e^{ks},
\end{equation}
where $k=1/\ell-i\omega/v_F$. For convergence of the integral $s_0<0$, and the limit $s_0\rightarrow-\infty$ is approached. When a boundary is hit, the solution should be constructed piecewise, and the boundary conditions applied in between.

\subsection{Boundary Conditions}\label{subsec-BC}
The boundary condition appropriate for specular scattering from a surface moving with velocity $\bm u$ is
\begin{equation}\label{spec bc2}
\psi_{\hat{\bm p}}=\psi_{\hat{\bm p}-2\hat{\bm n}(\hat{\bm n}\cdot\hat{\bm p})}+2p_{\rm F}(\hat{\bm n}\cdot\hat{\bm p})(\hat{\bm n}\cdot{\bm u}),
\end{equation}
where $\hat{\bm n}$ is the unit surface normal pointed to the liquid. In specular scattering, the incoming and reflected quasiparticle trajectories make the same angle to the surface normal on the wire. 
The boundary condition for diffusive scattering is that for outgoing quasiparticles ($\hat{\bm n}\cdot\hat{\bm p}_{\rm out}>0$)
\begin{eqnarray}\label{diff bc}
\psi_{\hat{\bm p}_{\rm out}}= -2\langle\hat{\bm n}\cdot\hat{\bm p}_{\rm in} \psi_{\hat{\bm p}_{\rm in}}\rangle_{\hat{\bm p}_{\rm in}}+p_{\rm F}(\hat{\bm p}_{\rm out}+\frac{2}{3}\hat{\bm n})\cdot{\bm u},
\end{eqnarray}
where $\langle\ldots\rangle_{\hat{\bm p}_{\rm in}}$ is an average over half of the unit sphere ($\hat{\bm n}\cdot\hat{\bm p}_{\rm in}<0$). In diffuse scattering the reflected (outgoing) quasiparticles on the surface of the wire are in equilibrium evaluated at the quasiparticle energy $\epsilon_{\bm p}=v_{\rm F}(p-p_{\rm F})+\delta\epsilon_{\hat{\bm p}}$. The distribution of reflected quasiparticles depends on the incoming quasiparticles only on the average.
We can also consider mixed boundary conditions where fraction $S$ of incoming quasiparticles is scattered specularly and fraction $1-S$ diffusely.
We note that in using the linearized boundary conditions (\ref{spec bc2}) and (\ref{diff bc}), the displacement of the wire leads to a second order correction, which is neglected, and thus the boundary conditions can be applied at the equilibrium location of the wire surface. 

At the container wall, one possiblity is to assume diffuse scattering, i.e.\ to use Eq. \eqref{diff bc} with ${\bm u}=0$.  An alternative, simpler boundary condition is an {\em absorbing wall}. This reflects no quasiparticles and is described by the condition $\psi_{\hat{\bm p}_{\rm out}}=0$, where $\hat{\bm n}\cdot\hat{\bm p}_{\rm out}>0$. Such a boundary condition could be a reasonable model for experimental cells that have walls made of sintered silver\cite{mart,Pentti09}.  The absorbing boundary condition is also a theoretical tool to suppress sound resonances (Sec.\ \ref{sec-res}). 

We demonstrate the generalization of Eq.\ (\ref{e.lbrel2x}) to include mixed boundary condition. By defining  $X_{\hat{\bm p}}(s)=\psi^{\rm le}_{\hat{\bm p}}(s)/\ell -i\omega\delta\epsilon_{\hat{\bm p}}(s)/v_{\rm F}$ we write
\begin{eqnarray}\label{spec.diff.ref.int}
\psi_{\hat{\bm p}}(s=0)=Sg_ce^{ks_c}+\left\{2Sp_{\rm F}(\hat{\bm n}_w\cdot\hat{\bm p})(\hat{\bm n}_w\cdot{\bm u})+(1-S)[p_{\rm F}(\hat{\bm p}+\textstyle{\frac23}\hat{\bm n}_w)\cdot{\bm u}+g_w] \right\}e^{ks_w}\nonumber\\
+\int_{s_w}^0 dsX_{\hat{\bm p}}(s)e^{ks} +S\int_{s_c}^{s_w} dsX_{\hat{\bm p}'}(s)e^{ks}
\end{eqnarray}
for a trajectory that is reflected from the wire (surface normal $\hat{\bm n}_w$) at $s=s_w$, and hits the container (surface normal $\hat{\bm n}_c$) at $s=s_c$.  In the latter integral the prime in direction $\hat{\bm p}'$ stands for the specularly reflected trajectory. We have defined two boundary condition terms \begin{eqnarray}
g_w(\bm r_w)=-2\hat{\bm n}_w\cdot\langle\hat{\bm p}_{\rm in} \psi_{\hat{\bm p}_{\rm in}}(\bm r_w)\rangle_{\hat{\bm p}_{\rm in}},\nonumber\\
g_c(\bm r_c)=-2\hat{\bm n}_c\cdot\langle\hat{\bm p}_{\rm in} \psi_{\hat{\bm p}_{\rm in}}(\bm r_c)\rangle_{\hat{\bm p}_{\rm in}},
\label{e.gwgc}\end{eqnarray}
which depend on the locations $\bm r_w$ and $\bm r_c$ on the wire and container surfaces, respectively.
For a trajectory with no collision with the wire, we get simply
\begin{eqnarray}\label{no-coll-int}
\psi_{\hat{\bm p}}(0)=g_ce^{ks_c}+\int_{s_c}^{0} dsX_{\hat{\bm p}}(s)e^{ks}.
\end{eqnarray}

\subsection{Force on the liquid}

The momentum flux tensor is\cite{VTres}
\begin{equation}\label{mom-flux}
\tensor\Pi=P^{(0)}\tensor1+\frac{\rho_s}{m_4}\delta\mu_4 \tensor 1 +3n_3\langle\hat{\bm p}\hat{\bm p}\psi_{\hat{\bm p}}\rangle_{\hat{\bm p}},
\end{equation}
where $P^{(0)}$ is  the equilibrium pressure. The superfluid density $\rho_s$ is defined as $\rho_s=m_4n_4-Dm^*n_3/(1+F_1/3)$, where $m_4$ is the mass of a $^4$He atom, $n_4$ the $^4$He number density, $D=1-(1+F_1/3)m_3/m^*$ and $n_3=p_{\rm F}^3/3\pi^2\hbar^3$  the number density of $^3$He. The force per area exerted by a surface element of the wire on the fluid is
$\hat{\bm n}\cdot\tensor\Pi$ evaluated at the wire surface $\bm r_w=a(\hat{\bm x}\cos\theta+\hat{\bm y}\sin\theta)$ . Integrating this over the perimeter of the wire  
gives the force
\begin{equation}\label{sym-force}
{\bm F}= \frac{a\rho_s}{m_4}\int_0^{2\pi}\delta\mu_4\hat{\bm n}\,d\theta+3an_3\int_0^{2\pi}\langle(\hat{\bm n}\cdot\hat{\bm p})\hat{\bm p}\psi_{\hat{\bm p}}\rangle_{\hat{\bm p}}\,d\theta.
\end{equation}
In linear theory both terms are proportional to $\bm u$ and thus this expression allows to determine $Z$ (\ref{zdef}). 

The two terms appearing in Eq.\ (\ref{sym-force}) can be interpreted as superfluid and normal  fluid contributions, respectively, and correspondingly $Z$ can be written as $Z=Z_s+Z_n$. We can calculate $Z_s$ as follows. Because the superfluid component is curl free, $\bm \nabla\times {\bm v}_s=0$, the fluid flow can be described as potential flow, ${\bm v}_s=\bm \nabla \chi$. The ideal fluid equation of motion\cite{VTres} $\partial {\bm v}_s/\partial t+\bm \nabla \delta\mu_4/m_4=0$ then gives $\delta\mu_4/m_4=-\partial\chi/\partial t$. For small frequencies $\omega \ll c/a$, $v_{\rm F}/a$ we can neglect the compressibility of the superfluid and the coupling to the normal component, and assume $\bm \nabla \cdot{\bm v}_s=0$, or $\nabla^2\chi=0$. This has to be solved using the boundary conditions $\hat{\bm n}\cdot{\bm v}_s=\hat{\bm n}\cdot{\bm u}$ on the wire surface and $\hat{\bm n}\cdot{\bm v}_s=0$ on the container walls.  Thus the problem reduces to solving the Laplace equation. The solutions in several geometries have been found, see 
Ref.\ \onlinecite{Brennen82}. The result is  
\begin{eqnarray}\label{e.he4vefs}
Z_{s}=-i\omega  \pi a^2 \rho_s \,G,
\end{eqnarray}
where the factor $G$ depends on the geometry of the container. For infinite fluid $G=1$, for the cylindrical container $G=(b^2+a^2)/(b^2-a^2)$, and for the slab container $G\approx1+\pi^2 a^2/12h^2$. 

In the limit $\ell\ll a$ the Fermi liquid theory reduces to the hydrodynamic theory (Sec.\ \ref{sec-hd}). In the extreme limit $\ell/a\rightarrow 0$ the viscosity is negligible, and the whole fluid behaves like an ideal fluid except thin boundary layers on the walls. In this limit $Z$ approaches
\begin{equation}\label{Z-ideal}
Z_{\rm ideal}=-i\omega  \pi a^2\rho \,G,
\end{equation}
where $\rho=m_4n_4+m_3n_3$ is the total density of the fluid. It is convenient to present $Z$ by its deviation from the ideal fluid behavior. Simultaneously, we define a dimensionless impedance $\tilde Z$ by writing
\begin{equation}\label{def-eff}
Z=Z_{\rm ideal} +an_3p_{\rm F}\tilde Z.
\end{equation}
Using the symmetry assumption $\bm F\parallel\bm u$ on equation \eqref{sym-force} and comparing to \eqref{def-eff} gives 
\begin{equation}\label{fct-gee}
\tilde Z=\frac{i\pi G}{1+F_1/3}\frac{a\omega}{v_{\rm F}}+\frac{3}{p_{\rm F}u}\int_0^{2\pi}\langle(\hat{\bm n}\cdot\hat{\bm p})(\hat{\bm p}\cdot\hat{\bm u})\psi_{\hat{\bm p}}\rangle_{\hat{\bm p}}\,d\theta,
\end{equation}
where the second term remains to be calculated numerically.
It follows from the equations above that  $\tilde Z$ depends on the dimensionless parameters $a\omega/v_{\rm F}(1+F_1/3)$, $\ell/a$, $S$, $F_0$, $F_1$, and  $b/a$ in the cylindrical container or $h/a$ in the slab container. Note that choosing $\Omega/(1+F_1/3)$ as an independent variable instead of $\Omega\equiv a\omega/v_{\rm F}$ is convenient because this combination frequently occurs in the hydrodynamic region, an example being the first term in Eq.\ (\ref{fct-gee}).

\section{Symmetry}\label{sec-sym}

The  problem stated in Sec.\ \ref{subsec-ma} has symmetries that can be used to simplify the numerical calculation. We need to consider scalar and vector functions of location $\bm r$. 
Because of the translational symmetry, there is no dependence on $z$, and therefore effectively $\bm r=\hat{\bm x}x+\hat{\bm y}y$, or using polar coordinates $\bm r=r(\hat{\bm x}\cos\theta+\hat{\bm y}\sin\theta)$. Because we are studying linear response, these functions depend linearly on the wire velocity $\bm u$. 
In the slab container  we have reflection symmetries in the planes perpendicular to both symmetry directions $x$ and $y$. Choosing $\hat{\bm u}=\hat{\bm x}$ we have for scalar functions $f$ and vector functions ${\bm f}=f_r\hat{\bm r}+f_\theta\hat{\bm \theta}$ the symmetries
\begin{eqnarray}\label{cb-symm-pol0}
f(r,\theta)&=&f(r,-\theta)=-f(r,\pi-\theta)=-f(r,\theta+\pi),\nonumber\\
f_r(r,\theta)&=&f_r(r,-\theta)=-f_r(r,\pi-\theta)=-f_r(r,\theta+\pi),\nonumber\\
f_\theta(r,\theta)&=&-f_\theta(r,-\theta)=f_\theta(r,\pi-\theta)=-f_\theta(r,\theta+\pi).
\end{eqnarray}

For the cylindrical container we argue as follows. Because of rotational symmetry in simultaneous rotations of $\bm r$ and $\bm u$, a scalar function has the form $f(\bm r,\bm u)=f(r,u,\theta)$, where $\theta$ denotes the relative plane angle and $r=|\bm r|$ and $u=|\bm u|$. Assuming reflection symmetry with respect to the $x$-axis $f(r,u,\theta)=f(r,u,-\theta)$. This allows to write $f(r, \hat{\bm r}\cdot \hat{\bm u}, u)$ or $f(r, \hat{\bm r}\cdot \bm u, u)$. Requiring $f$ to be linear in $\bm u$ limits this to the form \begin{eqnarray}
f(\bm r,\bm u)=\tilde f(r)\hat{\bm r}\cdot \bm u
\label{e.scsym}\end{eqnarray}
with some function $\tilde f(r)$.
Note that the symmetry (\ref{e.scsym}) often appears in solutions of the Laplace equation. Our derivation shows that this symmetry remains valid in the case of an arbitrary linear and isotropic equation.

Next, we consider a vector field $\bm f(\bm r,\bm u)=\hat{\bm r}f_r(\bm r,\bm u)+\hat{\bm \theta}f_\theta(\bm r,\bm u)$. Rotational symmetry limits the two functions to the forms $f_r(r,u,\theta)$ and $f_\theta(r,u,\theta)$. Assuming reflection symmetry  $f_r(r,u,\theta)=f_r(r,u,-\theta)$ allows to write this into the form $f_r(r, \hat{\bm r}\cdot \bm u, u)$. Assuming reflection symmetry  $f_\theta(r,u,-\theta)=f_\theta(r,u,\pi+\theta)$ allows to write this into the form $f_\theta(r, \hat{\bm \theta}\cdot \bm u, u)$. Finally, assuming linearity with respect to $\bm u$ implies 
\begin{eqnarray}
\bm f(\bm r,\bm u)=\hat{\bm r}\hat{\bm r}\cdot\bm u\tilde f_{r}(r) +\hat{\bm \theta}\hat{\bm \theta}\cdot\bm u\tilde f_{\theta}(r).
\label{e.scsymv}\end{eqnarray}

The results above are applied to the bulk angular averages 
\begin{equation}\label{beecee}
c(\bm r)=\langle \psi_{\hat{\bm p}}(\bm r)\rangle_{\hat{\bm p}},\;\;{\bm b}(\bm r)=3\langle \hat{\bm p}\psi_{\hat{\bm p}}(\bm r)\rangle_{\hat{\bm p}}
\end{equation}
and to the boundary-condition averages $g_w(\bm r_w)$ and $g_c(\bm r_c)$ (\ref{e.gwgc}).
The functions $c$ and $\bm b$ are needed to calculate $\delta\epsilon_{\hat{\bm p}}$ (\ref{e.deltae2})
and $\psi^{\rm le}_{\hat{\bm p}}$ (\ref{e.ledfd}).

\section{Hydrodynamic Limit}\label{sec-hd}

The Fermi-Bose liquid theory reduces to two-fluid hydrodynamic theory\cite{Kbook} in the limit of short mean-free-path \cite{khal,VTres}.
The normal fluid component is described by a Navier-Stokes equation where  the density is the normal fluid density $\rho_n=m^*n_3/(1+F_1/3)=\rho-\rho_s$ and the coefficient of viscosity $\eta=\frac15 n_3 p_{\rm F}\ell$. The diffusive boundary condition leads to no-slip boundary condition, which means that the fluid velocity at a wall equals the velocity of the wall. The specular boundary condition leads to perfect slip, where the transfer of transverse momentum between the liquid and the wall vanishes. Applied to the wire surface this means $\Pi_{r\theta}=0$. The hydrodynamic equations have analytic solution in some cases, which we discuss below. 

\begin{figure}[!tb]
\includegraphics[width=7cm]{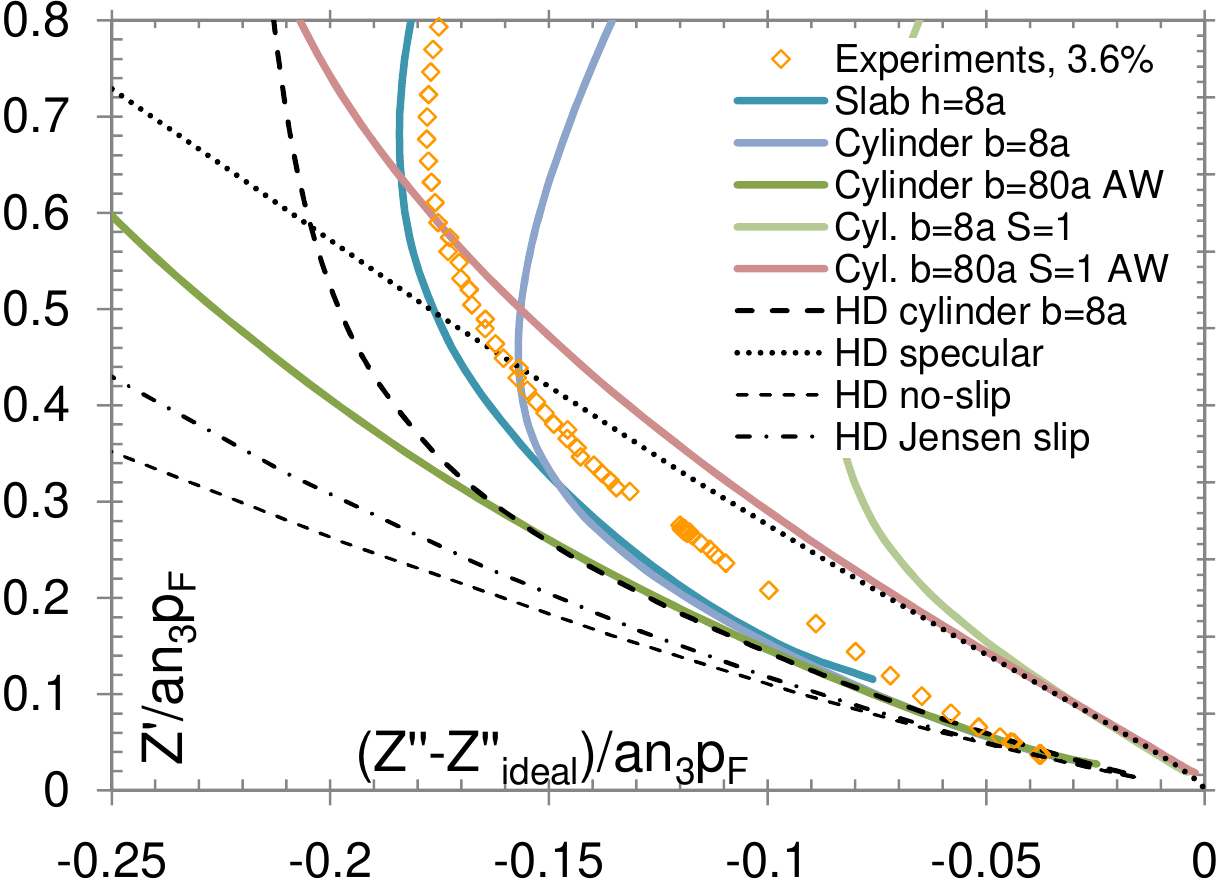}
\hspace{0.5cm}
\caption{The impedance $Z=Z'+iZ''$ in the region of short mean-free-paths, $\ell\lesssim a$. The curves are plotted using $\ell$ as parameter. The hydrodynamic results in unlimited fluid for diffusive wire (\ref{norm-sup-force}) and for specular wire (\ref{e.shz}) are shown by thin dashed and dotted lines, respectively. The slip correction to the former is shown by dash-dotted line\cite{HojgaardJensen80}. The hydrodynamic result for a diffuse wire in diffuse container of radius $b=8 a$ is shown by thick dashed line. Also shown are our numerical results: a large cylindrical container with a diffusive wire (green) and specular wire (pink) with absorbing walls, a small container, $b=8 a$, with a diffusive wire (light blue) and specular wire (light green) in diffusive container, and a slab of thickness $16 a$ (blue) with diffusive walls. The parameters are appropriate for a $3.6\%$ mixture, $a\omega/v_{\rm F}=0.0180$, $F_0=-0.25$, and $F_1=0.22$. Experimental data from Ref. \onlinecite{mart} is shown for reference (yellow points).}\label{Fig-hd-comp}
\end{figure}

In the low-frequency limit the normal and superfluid equations decouple and both components can be considered as incompressible. In this limit, the superfluid component $Z_s$ (\ref{e.he4vefs}) was found above. Analytic solutions for the normal component are known in the following cases. 
The no-slip case in unlimited liquid was solved by Stokes\cite{stokes}. The result is
\begin{equation}\label{norm-sup-force}
Z_n =- i\omega\pi a^2\rho_n [1 +\frac{4}{aq}\frac{H^{(1)'}_{0}(qa)}{H^{(1)}_{0}(qa)}].
\end{equation}
Here, $H_i^{(j)}(x)$ are the Hankel functions with complex argument, prime indicates derivative, $q=(1+i)/\delta$, and $\delta=\sqrt{2\eta/\rho_n\omega}$ is the viscous penetration depth.
The no-slip case in a cylindrical container was calculated by Carless, Hall, and Hook \cite{carl}.
The case of perfect slip in unlimited liquid was calculated by Bowley and Owers-Bradley \cite{bowley}, and the result is 
\begin{equation}\label{e.shz}
Z_n =- i\omega\pi a^2\rho_n[1 +\frac{8H^{(1)'}_{0}(qa)}{2aqH^{(1)}_{0}(qa)-a^2q^2H^{(1)'}_{0}(qa)}].
\end{equation}
The hydrodynamic results are illustrated in Fig.\ \ref{Fig-hd-comp}. 

The results (\ref{norm-sup-force}) and (\ref{e.shz}) illustrate the argument made in Sec.\ \ref{subsec-ma}
that $\tilde Z$ (\ref{def-eff}) is linear in $a \omega/v_{\rm F}$ but one should allow $a^2/\delta^2=\frac52a^2\omega/v_{\rm F}\ell(1+\frac13F_1)$ to have arbitrary magnitude. In the parameter space formed by $a \omega/v_{\rm F}$ and $\ell/a$, $\tilde Z$ is singular at the  point $(0,0)$. 

It is  possible to go beyond the low-frequency approximation at the expense of neglecting viscosity.
We study the case of the cylindrical container.
At frequency $\omega$ the velocity potential $\chi$ of ideal fluid should satisfy the wave equation, 
\begin{eqnarray}
\omega^2\chi+c^2\nabla^2\chi&=&0,
\end{eqnarray}
where $c$ is the sound velocity.
We make the ansatz
\begin{eqnarray}
\chi(r,\theta)=[AJ_1(kr)+BY_1(kr)]\cos\theta,
\end{eqnarray}
where $J_1$ and $Y_1$ are Bessel functions and here $k=\omega/c$.
The radial velocity
has to equal $u\cos\theta$ at $r=a$ and vanish at $r=b$. This gives conditions from which $A$ and $B$ can be solved. Calculating the force similarly as for the superfluid component in Eq.\ (\ref{sym-force}), we get
\begin{eqnarray}
Z_{\rm ideal}=\frac{i\pi a\rho\omega}{k}\frac{J_1(ka)Y_1'(kb)-Y_1(ka)J_1'(kb)}{J_1'(ka)Y_1'(kb)-Y_1'(ka)J_1'(kb)},
\label{e.zserosre}\end{eqnarray}
where  $\rho$ is the density of the fluid.
To first order in $\omega$ this reduces to the low-frequency result $Z_{\rm ideal}$ (\ref{Z-ideal}).
Sound resonances are found at frequencies where the denominator of $Z_{\rm ideal}$
(\ref{e.zserosre}) vanishes. For large $b/a$ this condition reduces to zeros of $J_1'(kb)$. The lowest zero appears at $kb=1.84$.

The two-fluid hydrodynamics allows two sound modes. The first sound is in-phase motion of the normal and superfluid components. The corresponding sound velocity is close to the sound velocity of pure liquid $^4$He, which is relatively high, and is not of interest here. Second sound is the counter motion of the normal and superfluid components. In $^3$He-$^4$He mixture it is essentially a compressional mode of the normal fluid as the superfluid motion causes only a small correction to the sound velocity because of smallness of $n_3/n_4$ [\onlinecite{khal}]. In the present approximation (Sec.\ \ref{subsec-ma}) the sound velocity is $v_{\rm F}\sqrt{(1+F_1/3)(1+F_0)/3}\approx 0.4 v_{\rm F}$ in the absence of dissipation. This result will be applied below to identify resonances of second sound.

An extension of the hydrodynamic theory to slightly longer mean-free-path is known as slip theory\cite{EinzelParpia97}. The leading corrections appear in the boundary conditions. Considering the diffusive boundary condition (\ref{diff bc}), the outgoing (scattered from the wall) quasiparticles are in equilibrium with the wall, but the incoming quasiparticles generally are not. In the case of short $\ell$ these populations get rapidly mixed and therefore  also the incoming quasiparticles are in equilibrium with the wall, which leads to the no-slip boundary condition. With increasing $\ell$ this is no more the case. In slip theory one uses bulk hydrodynamic theory together with boundary conditions that assume the velocity of the fluid to extrapolate to the wall velocity the slip length $\zeta$ behind the wall.
Microscopic calculation of $\zeta$ for degenerate fermions on a planar wall was made by Jensen \emph{et al.}\cite{HojgaardJensen80}. The generalization of the slip boundary condition to curved surfaces and partial specularity is discussed in Refs.\ \onlinecite{Einzel90,bowley,Perisanu06}.
The slip increases the validity range of the hydrodynamic theory to slightly larger $\ell$, see Fig.\ \ref{Fig-hd-comp}. Several extrapolations of the slip theory for vibrating wires have been suggested, see  Carless, Hall and Hook\cite{carl}, Guénault \emph{et al.}\cite{gue1}, and more recently by Bowley and Owers-Bradley\cite{bowley} together with Perisanu and Vermeulen\cite{Perisanu06}. None of these have attempted to include the effect of the container or the Fermi-liquid interactions.

\section{Ballistic Limit}\label{sec-bal}

At the lowest temperatures one enters the ballistic regime, where the mean-free-path of quasiparticles becomes large compared to the experimental dimensions. In the extreme case one can neglect  the collision term in the kinetic equation (\ref{e.lbrel2}). In this ballistic limit the quasiparticles still interact through the Fermi-liquid interactions, which appear through $\delta\epsilon_{\hat{\bm p}}$. In spite of this, some results can be obtained analytically\cite{bowley,VT,VT09}.

If, in addition to the collision term, one also neglects the time dependence, the  kinetic equation (\ref{e.lbrel2}) reduces to $\hat{\bm p}\cdot\nabla\psi_{\hat{\bm p}}({\bm r})=0$. This implies that $\psi_{\hat{\bm p}}$ is constant along trajectories, and changes only when the trajectory hits a wall. If the fluid far from the wire is in equilibrium,  one finds $\psi_{\hat{\bm p}}=0$ on incoming trajectories. The distribution for outgoing trajectories is then obtained from the boundary conditions (\ref{spec bc2}) or (\ref{diff bc}). The impedance is obtained by integrating over the wire surface, and it gives\cite{bowley,VT}
\begin{eqnarray}\label{simple force}
 Z_{\rm diff}&=&\frac{43\pi}{48}a n_{3} p_{\rm F},\nonumber\\
 Z_{\rm spec}&=&\frac{3\pi}{4}an_{3}p_{\rm F}
\end{eqnarray}
corresponding to the diffusive and specular boundary conditions. These are purely dissipative since our assumption of $\omega\rightarrow 0$. These results have been generalized to the cylindrical container by taking into account the reflection of quasiparticles from the container wall back to the wire in Ref.\ \onlinecite{VT09}.

At finite frequency the main complication arises from the interaction term $\delta\epsilon_{\hat{\bm p}}$ in the kinetic equation (\ref{e.lbrel2}). Neglecting this term it is possible to solve the ballistic limit at arbitrary frequency in the cylindrical container. We assume diffusive boundary condition on both the wire and the container. Using definitions (\ref{e.gwgc}) and the symmetry (\ref{e.scsym}), the boundary conditions (\ref{diff bc}) can be written as
\begin{eqnarray}
\psi_{\hat{\bm p}_{\rm w, out}}(\bm r_w)&=&\tilde g_w\hat{\bm n}\cdot{\bm u}+p_{\rm F}(\hat{\bm p}_{\rm w, out}+\frac{2}{3}\hat{\bm n})\cdot{\bm u},\\
\psi_{\hat{\bm p}_{\rm c, out}}(\bm r_c)&=&\tilde g_c\hat{\bm n}_c\cdot{\bm u},
\end{eqnarray}
where $\tilde g_w$ and $\tilde g_c$ are constants. Tracing the different trajectorices one can now calculate the averages appearing in definitions (\ref{e.gwgc}). We get $\tilde g_w=D \tilde g_c$ and
\begin{eqnarray}\label{bal-int-def}
\tilde g_c=A\tilde g_c+B(\tilde g_w+\frac23 p_{\rm F})+\frac23Cp_{\rm F}.
\end{eqnarray}
The coefficients $A$, $B$, $C$, and $D$ are expressed as integrals
\begin{eqnarray}\label{int-Aa}
A&=&\frac{2}{\pi}\int_0^\pi d\zeta\sin^2\zeta\int_{\arcsin (a/b)}^{\pi/2}d\gamma \cos\gamma(1-2\cos^2\gamma)\exp\left(\frac{2i\omega b\cos\gamma}{v_{\rm F}\sin\zeta}\right),\nonumber\\
B&=&\frac ab\frac{2}{\pi}\int_0^\pi d\zeta\sin^2\zeta\int_{0}^{1}dx
\left(\frac {a}{b}x^2+\sqrt{1- \frac{a^2}{b^2}x^2}\sqrt{1-x^2}\right)\exp\left(\frac{i\omega (\sqrt{b^2-a^2x^2}-a\sqrt{1-x^2})}{v_{\rm F}\sin\zeta}\right)=\frac ab D,\nonumber\\
C&=&\frac ab\frac{3}{\pi}\int_0^\pi d\zeta\sin^3\zeta\int_{0}^{1}dx
\sqrt{1- \frac{a^2}{b^2}x^2}\exp\left(\frac{i\omega (\sqrt{b^2-a^2x^2}-a\sqrt{1-x^2})}{v_{\rm F}\sin\zeta}\right),
\nonumber\\
D&=&\frac{2}{\pi}\int_0^\pi d\zeta\sin^2\zeta\int_{0}^{1}dx
\left(\frac {a}{b}x^2+\sqrt{1-x^2}\sqrt{1- \frac{a^2}{b^2}x^2}\right)\exp\left(\frac{i\omega (\sqrt{b^2-a^2x^2}-a\sqrt{1-x^2})}{v_{\rm F}\sin\zeta}\right),
\label{e.dintom}\end{eqnarray}
where $\zeta$ is the angle between the trajectory and the cylinder axis. 
We can now solve (\ref{bal-int-def}) and get
\begin{eqnarray}
\tilde g_c=\frac{\frac23(B+C)p_{\rm F}}{ 1-A-BD}.
\label{e.gcw}\end{eqnarray}
Continuation to calculate the force gives 
\begin{eqnarray}
Z_n=\pi an_3\left(\frac{43 p_{\rm F}}{48}+\frac 12 \tilde g_w+\frac{b}{2a}C\tilde g_c\right)
=\pi an_3p_{\rm F}\left(\frac{43 }{48}+\frac {a}{3b}\frac{(D+\frac{b}{a}C)^2}{1-A-\frac{a}{b}D^2} \right).
\label{e.foraabd}\end{eqnarray}
In the special case $\omega=0$ the integrals (\ref{e.dintom}) reduce to
\begin{eqnarray}
A&=&-\frac13+\frac ab-\frac23\frac {a^3}{b^3},\nonumber\\
C
&=&\frac2{\pi}\left(\frac ab\sqrt{1-\frac{a^2}{b^2}}+\arcsin\frac ab\right),\nonumber\\
D&=&\int_{0}^{1}dx \left(\frac {a}{b}x^2+\sqrt{1-x^2}\sqrt{1- \frac{a^2}{b^2}x^2}\right).
\label{e.dint}\end{eqnarray}
where $D$ still needs to be calculated numerically.
These reproduce the result given in Ref.\ \onlinecite{VT09}.

\begin{figure}[!tb]
\centerline{
\includegraphics[width=0.32\linewidth]{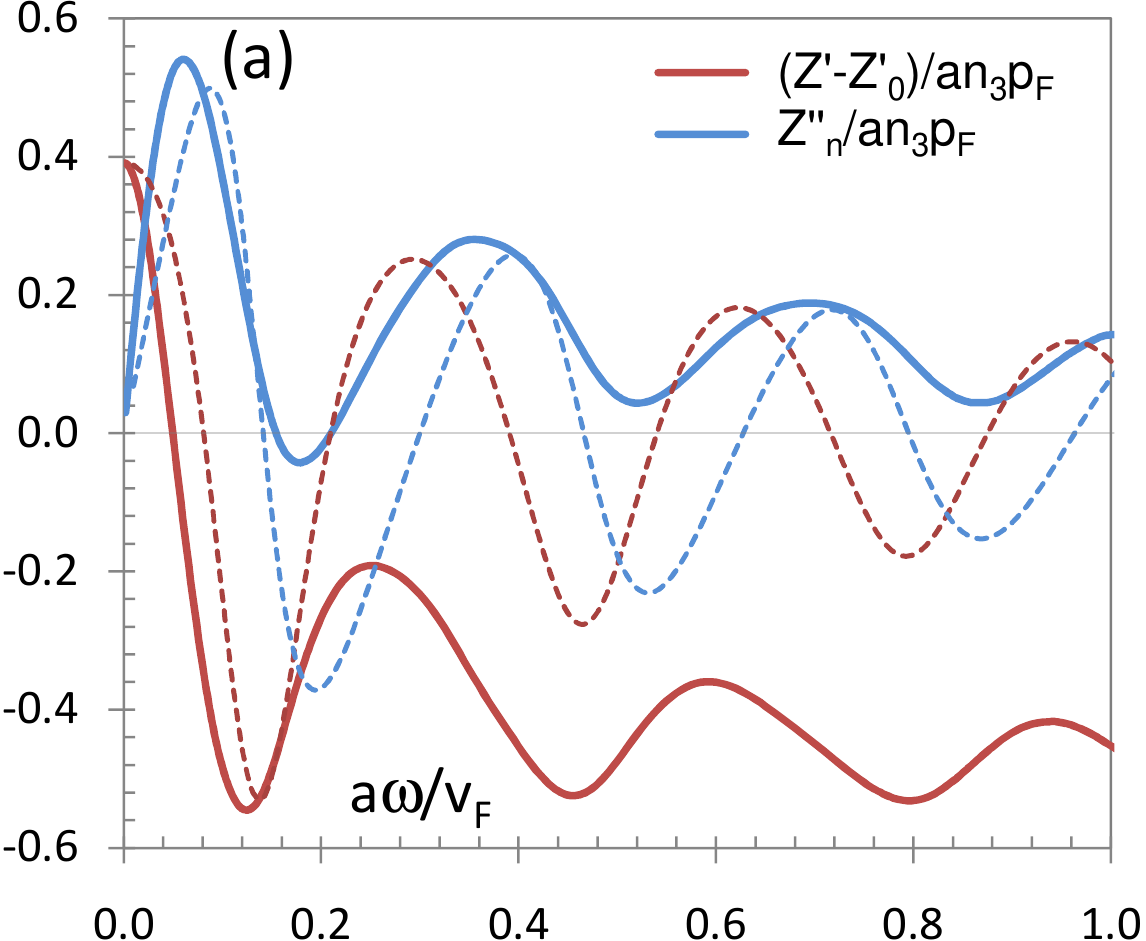}
\includegraphics[width=0.27\linewidth]{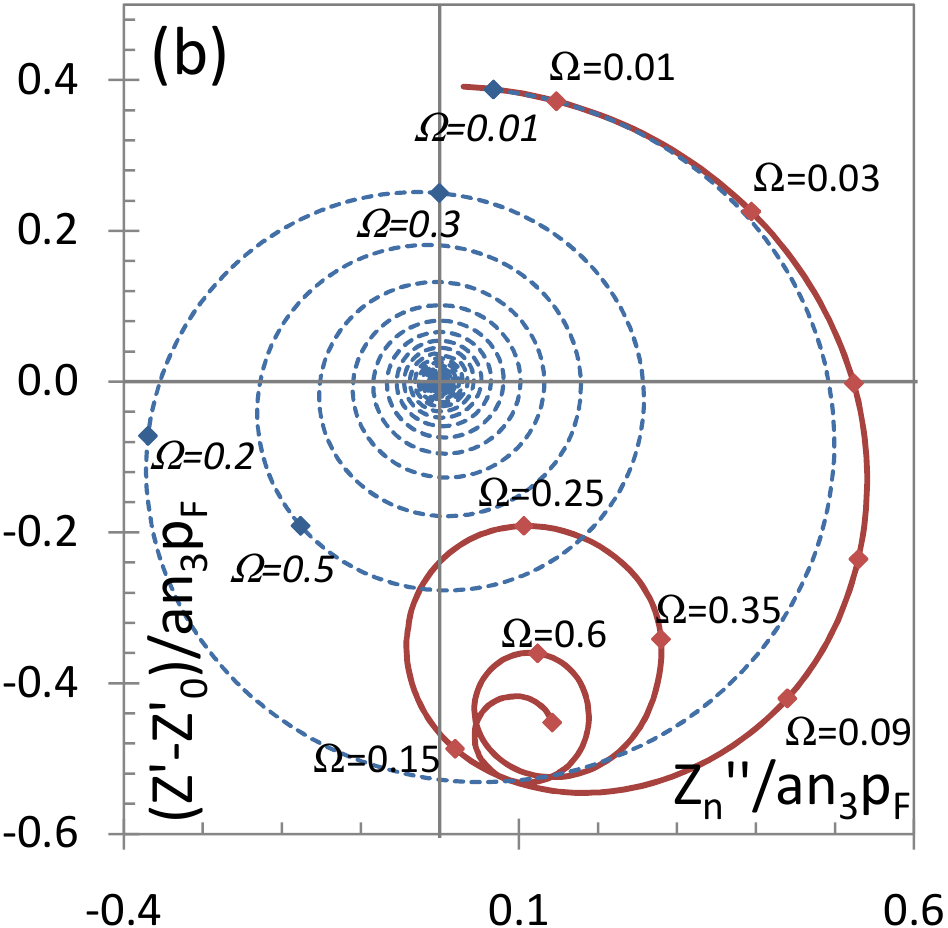}
}\caption{The normal-fluid impedance $Z_n=Z'+iZ_n''$ in the ballistic limit in the cylindrical container of radius $b=10a$. The impedance $Z_n$ is plotted relative to its static value $Z_0=(43\pi/48)an_3p_{\rm F}$ (\ref{simple force}) in unlimited fluid, and the total impedance $Z$ is obtained by adding $Z_s$ (\ref{e.he4vefs}). The solid lines are full numerical result, the dashed lines the approximation (\ref{e.foraabd}), and the difference between them is the Landau force \cite{VTletter}. (a) The real (red) and imaginary (blue) parts of impedance as a function of $\Omega=a\omega/v_{\rm F}$. (b) Parametric plot of $Z_n$ in complex plane with $\Omega$ as a parameter. 
In both panels diffusive boundary conditions are used and the full calculation uses $F_0=-0.44$ and $F_1=0.449$. 
}
\label{f.balom}
\end{figure}

The result (\ref{e.foraabd}) is illustrated in Fig.\ \ref{f.balom}. We see that $Z_n$ oscillates around its static value approximately like  the exponential $\propto\exp(2i\omega b/v_{\rm F})$.  The reason is that the quasiparticles excited by the wire are reflected from the container wall back to the wire, but delayed by time $\approx 2b/v_{\rm F}$. The first constructive interference (after the zeroth one at $\omega=0$) corresponds to $\omega\approx\pi v_{\rm F}/b$. This is approximately a factor 4 larger than the first second-sound resonance frequency in the hydrodynamic limit (Sec.\ \ref{sec-hd}). The difference arises because the second sound velocity is smaller than the quasiparticle velocity $v_{\rm F}$, and because the first zero of $J_1'(kb)$ is at $kb=1.84$, which is less than  $\pi$. Second sound is strongly damped in the ballistic regime because of excitation of quasiparticles, known as Landau damping.

The results above could be generalized to include the Fermi-liquid interactions. However, the integrals get very cumbersome even in the case of unlimited liquid. Instead of partial analytic solution, we therefore prefer the full numerical solution in the following.

\section{Numerical method}\label{sec-num}

We solve the quasiparticle distribution $\psi_{\hat{\bm p}}({\bm r})$ in a discrete grid around the wire and for discrete momentum directions $\hat{\bm p}$. The location ${\bm r}_n=\hat{\bm  x}r\cos\theta +\hat{\bm y}r\sin\theta$ can be parameterized with cylindrical coordinates $r$ and $\theta$, where we have fixed $\hat{\bm x}=\hat{\bm u}$. The momentum direction $\hat{\bm p}_{jl}=\sin\zeta_j(\cos\beta_l\hat{\bm x}+\sin\beta_l\hat{\bm y})+\cos\zeta_j\hat{\bm z}$ is parameterized by angles $\zeta_j$ and $\beta_l$. For the cylindrical container we use a grid with exponentially increasing spacing $\Delta r$ in the radial direction, and a fixed spacing $\Delta\theta$. In the slab container, we use a cylindrical lattice near the wire, and rectangular lattice near the container walls, with partial overlap between the lattices. In $y$-direction the length of the slab is increased until the results converge (absorbing walls are used).

The quasiparticle distribution $\psi_{njl}\equiv\psi_{\hat{\bm p}_{jl}}({\bm r}_n)$ for each lattice point $n$ and for each discrete trajectory direction $\hat{\bm p}_{jl}$ can be written, using Eq.\ \eqref{spec.diff.ref.int}, as the sum  
\begin{eqnarray}\label{full-psi-njk}
\psi_{njl}=I_{njl}+B_{njl}+S\,H_{njl}+(1-S)Q_{njl},
\end{eqnarray}
where $I_{njl}=\int_{s_w}^0ds\,X_{\hat{\bm p}_{jl}}(s)e^{ks}+S\int_{s_c}^{s_w}ds\,X_{\hat{\bm p}'_{jl}}(s)e^{ks}$ is the integral part, $B_{njl}=[2Sp_{\rm F}(\hat{\bm n}\cdot\hat{\bm p})(\hat{\bm n}\cdot{\bm u})+(1-S)p_{\rm F}(\hat{\bm p}+\frac23\hat{\bm n})\cdot{\bm u}]e^{ks_w}$ is the $\psi$-independent part of the boundary condition, $H_{njl}$ describes the $\psi$-dependent part of the boundary condition on the container, and $Q_{njl}$ describes the boundary condition on the wire. The  integral is calculated from ${\bm r}_n$ along the trajectory $\hat{\bm p}_{jl}$, $|s_w|$ is the distance to the wire (if there is a collision with the wire at point ${\bm r}_w$), and $|s_c|$ is the distance at which the container wall is hit, at point ${\bm r}_c$. If there is no collision with the wire, we get from 
Eq.\ \eqref{no-coll-int} $\psi_{njl}=I_{njl}+H_{njl}$, where $I_{njl} =\int_{s_c}^0ds\,X_{\hat{\bm p}_{jl}}(s)e^{ks}$. We write $g_{c,t}=g_c({\bm r}_t)$ and $g_{w,t}=g_w({\bm r}_t)$ at each lattice point ${\bm r}_t$ on the surfaces, and define
\begin{eqnarray}
H_{njl}&=&\sum_t\, H_{ntjl}\,g_{c,t},\nonumber\\Q_{njl}&=&\sum_t\, Q_{ntjl}\,g_{w,t},
\end{eqnarray}
where the sums are over all lattice points on the corresponding surface, $H_{ntjl}=w_t e^{ks_c}$ and $Q_{ntjl}=w_t e^{ks_w}$. The weight factors $w_t$ are obtained by interpolating from the lattice points closest to ${\bm r}_c$ or ${\bm r}_w$ on the surfaces. In the following the averages \eqref{beecee} are denoted by $\Psi_n^{(1)}=c_n$, $\Psi_n^{(2)}=b_{r,\,n}$, and $\Psi_n^{(3)}=b_{\theta,\,n}$. Now, the integral part $I_{njl}$ can be conveniently presented as the sum
\begin{eqnarray}
I_{njl}=\sum_{h=1}^3\,\sum_{m}\, a_{nmjl}^{(h)}\,\Psi_m^{(h)},
\end{eqnarray}
where the coefficients $a_{nmjl}^{(h)}$ give the weights with which the equilibrium distribution at other lattice points ${\bm r}_m$ affects $\psi_{njl}$. The coefficients are calculated by numerical integration, using a modified Simpson's rule, which takes the exponential term $\exp(ks)$ into account exactly. It is necessary to use interpolation between lattice points. Taking the numerical average over the trajectories, we can write the averages \eqref{beecee} as a sum
\begin{eqnarray}\label{le-psi-njk}
\Psi_n^{(i)}=\sum_{m,h}d_{nm}^{ih}\Psi_m^{(h)}+B_n^{(i)}+\sum_tH_{nt}^{(i)}g_{c,t}+\sum_tQ_{nt}^{(i)}g_{w,t},
\end{eqnarray}
where
\begin{eqnarray}\label{aver-coeff}
d_{nm}^{1h}&=&\langle a_{nmjl}^{(h)}\rangle_{\hat{\bm p}_{jl}},\nonumber\\d_{nm}^{2h}&=&3\hat{\bm r}_n\cdot\langle \hat{\bm p}_{jl} a_{nmjl}^{(h)}\rangle_{\hat{\bm p}_{jl}},\nonumber\\d_{nm}^{3h}&=&3\hat{\bm \theta}_n\cdot\langle \hat{\bm p}_{jl} a_{nmjl}^{(h)}\rangle_{\hat{\bm p}_{jl}},
\end{eqnarray}
and similar averaging gives $B_n^{(i)}$, $H_{nt}^{(i)}$, and $Q_{nt}^{(i)}$. The diffusive boundary condition terms can be written as sums over lattice points $m$ as well,
\begin{eqnarray}
g_{c,t}&=&\sum_{m,h} g_{c,tm}^{(h)}\Psi^{(h)}_m+ B_{c,t}+\sum_s H_{c,ts}g_{c,s}+\sum_s Q_{c,ts}g_{w,s},\nonumber\\ g_{w,t}&=&\sum_{m,h} g_{w,tm}^{(h)}\Psi^{(h)}_m+ \sum_s H_{w,ts}g_{c,s},
\end{eqnarray}
where the new coefficients $g_{c,tm}^{(h)}$, $B_{c,t}$, $H_{c,ts}$, $Q_{c,ts}$, $g_{w,tm}^{(h)}$, and $H_{w,ts}$ are similar to the terms above, but the averages are taken over the incoming trajectories only. Once the coefficients have been calculated, we can solve the averages \eqref{beecee} from the matrix equation
\begin{eqnarray}\label{matrix-eq}
\Psi=D\Psi+B\Leftrightarrow \Psi=(I-D)^{-1}B.
\end{eqnarray}
If the total number of lattice points is $N$,  the number of points on container surface is $V$, and the number of points on wire surface is $T$, then $\Psi$ is a $3N+V+T$ component vector $\Psi=(\Psi_1^{(1)},\Psi_2^{(1)},\ldots,\Psi_N^{(3)},g_{c,1},\ldots,g_{c,V},g_{w,1},\ldots,g_{w,T})^T$, $B=(B_1^{(1)},B_2^{(1)},\ldots,B_N^{(3)},B_{c,1},\ldots,B_{c,V},0,\ldots,0)^T$, and the $(3N+V+T)\times(3N+V+T)$ matrix $D$ contains the terms $d_{nm}^{ih}$, $H_{nt}^{(i)}$, $Q_{nt}^{(i)}$, $g_{c,tm}^{(h)}$, $H_{c,ts}$, $Q_{c,ts}$, $g_{w,tm}^{(h)}$, and $H_{w,ts}$ in appropriate order.

When the averages are known, we can calculate the full distribution $\psi_{njl}$ on the wire surface using Eq.\ \eqref{full-psi-njk}. The force exerted by the wire on the fluid is then readily calculated from 
Eq.\ \eqref{fct-gee}.  If cylindrical symmetry is assumed, the treatment is similar, but since the dependence of the averages on $\theta$ is of a simple form [(\ref{e.scsym}) and  (\ref{e.scsymv})], the calculation becomes effectively one-dimensional.

\section{Parameter values}\label{sec-pv}
In order to relate the numerical calculations to experiments we need to consider the parameters that characterize $^3$He-$^4$He mixtures and the experimental setup. 

\begin{table}
\begin{tabular}{c||c|c|c|c||c|c|c|c|c|c|c||c|c}
\hline
\hline
  &$\ $ 1.8\%&$\ $ 3.6\%&$\ $ 5.6\%&$\ $ 6.6\%&$\ $ 7.0\%&$\ $ 9.5\%& $\ \ \ $&$\ $ 1.8\% &$\ $ 3.6\%&$\ $ 5.6\%&$\ $ 6.6\%&$\ $ 7.0\%&$\ $ 9.5\%\\
  \hline
 $m^*/m_3$            & 2.26  &  2.31  &  2.34  &  2.35  &  2.60  &  2.63   & $\ \ \ $ & 2.45  &  2.50  &  2.53  &  2.54  &  2.85  &  2.88 \\
 $v_{\rm F}$(m/s)     & 21.10 &  25.97 &  29.60 &  31.07 &  29.56 &  32.36  & $\ \ \ $ & 19.46 &  24.00 &  27.38 &  28.75 &  26.97 &  29.55 \\
 $an_3p_{\rm F}$ (kg/ms) & 0.0058& 0.0145 & 0.0259 & 0.0321 & 0.0394 & 0.0589 & $\ \ \ $ & 0.0058& 0.0145 & 0.0259 & 0.0321 & 0.0394 & 0.0589  \\
 $F_1$                & 0.151 &  0.219 &  0.266 &  0.284 &  0.269 &  0.301  & $\ \ \ $ & 0.138 &  0.201 &  0.245 &  0.261&  0.243 &  0.272 \\
 $a\omega/v_{\rm F}$, wire 1& 0.0222& 0.0180 & 0.0158 & 0.0151 & 0.0159 & 0.0145  & $\ \ \ $ & 0.0241& 0.0195 & 0.0171 & 0.0163 & 0.0174 & 0.0159 \\
 $a\omega/v_{\rm F}$, wire 2&    -  &     -  &     -  & 0.0233 & 0.0245 & 0.0224  & $\ \ \ $ &    -  &     -  &     -  & 0.0252 & 0.0268 & 0.0245 \\
  \hline
  \hline
\end{tabular}
\caption{Parameter values used in calculations at different concentrations $x_3$. The first four concentrations are at saturated vapor pressure, the last two at $P=10$ atm. We use $\alpha=0.284$ at saturated vapor pressure and $\alpha=0.207$ at $P=10$ atm\cite{wat}. The frequency parameter $\Omega=a\omega/v_{\rm F}$ is calculated for vacuum frequency $f=1202.85$ Hz (wire 1) and $f=1857.7$ Hz (wire 2). On the left we have used the hydrodynamic mass $m_H/m_3=2.15$ as a starting point, and on the right $m_H/m_3=2.34$ (see text).}\label{tab4}
\end{table}

The molar volume of a mixture of molar concentration $x_3=N_3/(N_3+N_4)$ at temperature $T$ and pressure $P$ is\cite{wat}
\begin{eqnarray}
V_m(x_3,P,T)=V_4(P,T)[1+\alpha(x_3,P,T)x_3],
\end{eqnarray}
where $V_4(P,T)$ is the molar volume of pure $^4$He and $\alpha$ is the so-called BBP parameter\cite{BBP}. For the molar volume of pure $^4$He, we use the data of Tanaka \emph{et al.}\cite{tanaka}, and for $\alpha(P)$ at zero temperature we use the results of Watson \emph{et al.}\cite{wat}. The number density of the $^3$He component is obtained from
\begin{equation}
n_3=\frac{N_Ax_3 }{V_4(1+\alpha x_3)},
\end{equation}
where $N_A$ is the Avogadro number. From $n_3$ we can calculate the Fermi momentum
\begin{equation}
p_{\rm F}=\sqrt[3]{3\pi^2\hbar^3n_3}.
\end{equation}
If we know the $^3$He quasiparticle effective mass $m^*$, we can also calculate the Fermi velocity $v_{\rm F}=p_{\rm F}/m^*$, which is needed in calculating the dimensionless frequency parameter $\Omega=a\omega/v_{\rm F}$. The effective mass $m^*$ is calculated using the interpolation formula by Krotscheck \emph{et al.}\cite{krot,krot2}. As a starting point, the zero concentration limit effective mass (so called hydrodynamic mass) $m_H$ is needed. We have used two alternative values: $m_H/m_3=2.15$ based on extrapolation of the data by Simons et al.\cite{simons} following Krotscheck \emph{et al.}\cite{krot}, and $m_H/m_3=2.34$\cite{anderson,BBP2,hsu} at zero pressure. 
Here $m_3$ is the mass of a $^3$He atom. At higher pressure $P=10$ atm we have used $m_H/m_3=2.39$ and $m_H/m_3=2.64$, correspondingly. Numerical values are given in Table \ref{tab4}. For $F_1$ we use the approximative\cite{BaymPethick} formula $F_1=3(m^*/m_H-1)$. For $F_0$ we use a fit of the form $F_0=Ax_3^{1/2}+Bx_3^{1/3}$ to two sets of data presented in Ref.\ \onlinecite{corruccini}, see Table \ref{tab5}.

\begin{table}
\begin{tabular}{c||c|c|c|c||c|c}
\hline
\hline
&$\ $ 1.8\%&$\ $ 3.6\%&$\ $ 5.6\%&$\ $ 6.6\%&$\ $ 7.0\%&$\ $ 9.5\% \\
 \hline
$F_0$, Ref.\ [\onlinecite{landau,landau2}]& -0.17& -0.25 & -0.33 & -0.36 & -0.38 & -0.47  \\
$F_0$, Ref.\  [\onlinecite{murdock}]& -0.12& -0.21 & -0.28 & -0.32 & -0.24 & -0.29  \\
 \hline
 \hline
\end{tabular}
\caption{$F_0$ at different concentrations based on data presented in Ref.\ [\onlinecite{corruccini}]. The first line gives results from a fit to the osmotic pressure data of Landau \emph{et al.}\cite{landau,landau2}, and the second line gives a fit to the second sound velocity data by Murdock and Corrucini\cite{murdock}.
The first four concentrations are at saturated vapor pressure, the last two at $P=10$ atm.
}\label{tab5}
\end{table}

We present $Z$ \eqref{def-eff} using the dimensionless $\tilde Z$ in order to remove the strong dependence $\propto n_3^{4/3}$  on the $^3$He density. The experimental results\cite{mart,gue1} are given as $(f_0,\Delta f)$ pairs. In order to transform them to $\tilde Z$ we use Eqs.\ \eqref{Z-freq} and \eqref{def-eff}, or 
\begin{eqnarray}
\tilde Z'&=&\frac{2\pi^2a^2\rho_w}{an_{3}p_{\rm F}}\Delta f,\nonumber\\
\tilde Z''&=&\frac{4\pi^2a^2\rho_w}{an_{3}p_{\rm F}}[(1+\frac12G\frac{\rho}{\rho_w})f_0-f_{\rm vac}].
\label{e.mesxtozconv}\end{eqnarray}
Several vibrating wire measurements have been done in $^3$He-$^4$He mixtures  \cite{carl,gue1,konig,Perisanu06}. Here we study the data of Martikainen \emph{et al.}\cite{mart,Pentti09}, which extends deepest into  the ballistic regime. The parameters needed for the conversion (\ref{e.mesxtozconv}) are the wire radius $a=62$ $\mu$m, the distance between the container walls $2h=16 a$, the vacuum frequencies $f_{\rm vac}=1202.85$ Hz for wire 1 and $f_{\rm vac}=1857.7$ Hz for wire 2, the density of the wire $\rho_w=16700$ kg/m$^3$ and the concentration-dependent densities as explained above. The vacuum frequencies $f_{\rm vac}$ were measured before filling the cell. The measurements with filled cell show small deviation from the predicted high temperature behavior as if the vacuum frequency had some variation during the measurements. We have compensated this by adjusting $f_{\rm vac}$ for each concentration separately in the range $\pm 0.2$ Hz so that at high temperatures $\tilde Z$ extrapolates to zero.

\section{Results and Discussion}\label{sec-res}
In this section we present the results of numerical calculations for the full scale of the mean-free-path $\ell$. 
In order to show the dependence of $Z$ on different parameters, we select a ``basic set'' of parameters: $\Omega/(1+F_1/3)=0.0145$ (where  $\Omega\equiv a\omega/v_{\rm F}$), the cylindrical container with $b=10 a$, $S=0$, $F_0=-0.33$, $F_1=0.266$, and diffusely scattering container walls. Then we vary each of the parameters $\Omega/(1+F_1/3)$, $b/a$, $S$, $F_0$, and $F_1$ separately while keeping the others fixed at the basic set values. The real and imaginary parts of the impedance $Z$ are interpreted as dissipation and shift of resonance frequency, see Eq.\ (\ref{Z-freq}).

\begin{figure}[!tb]
\centerline{
\includegraphics[width=0.32\linewidth]{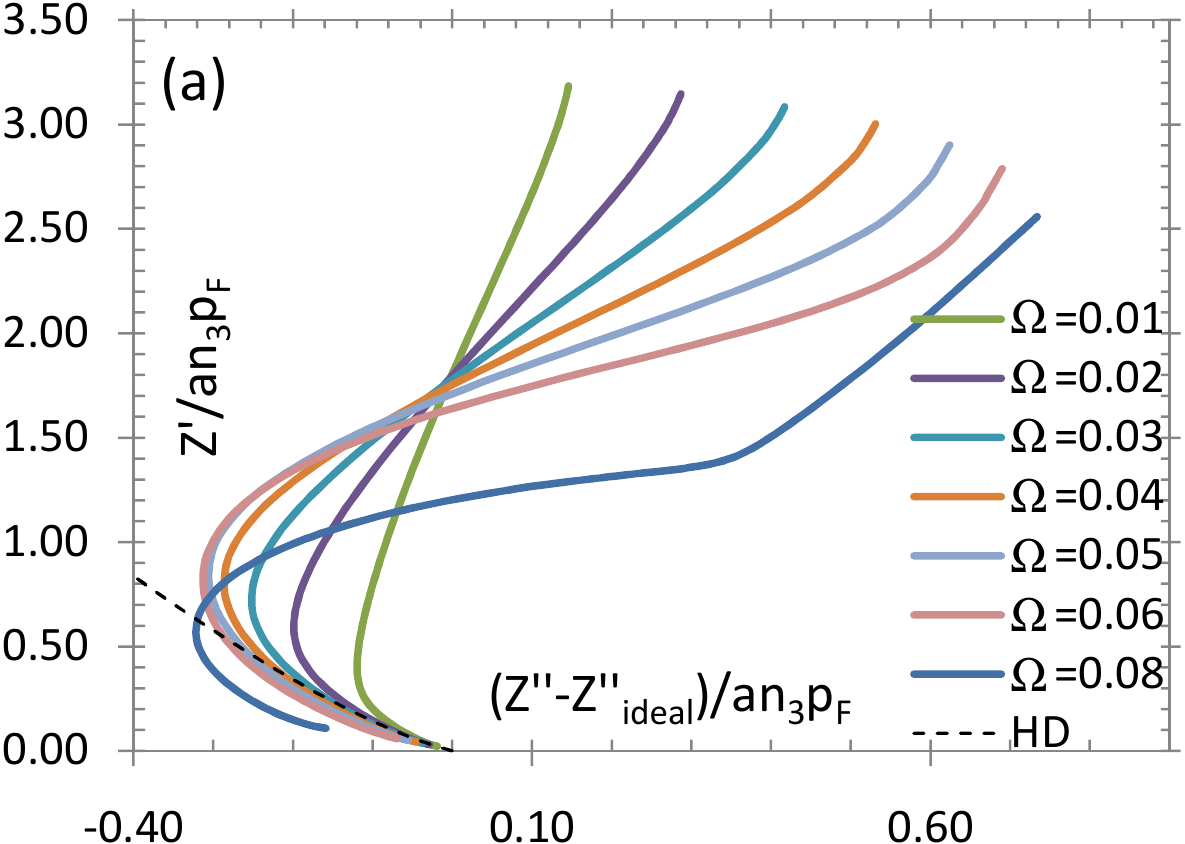}
\includegraphics[width=0.32\linewidth]{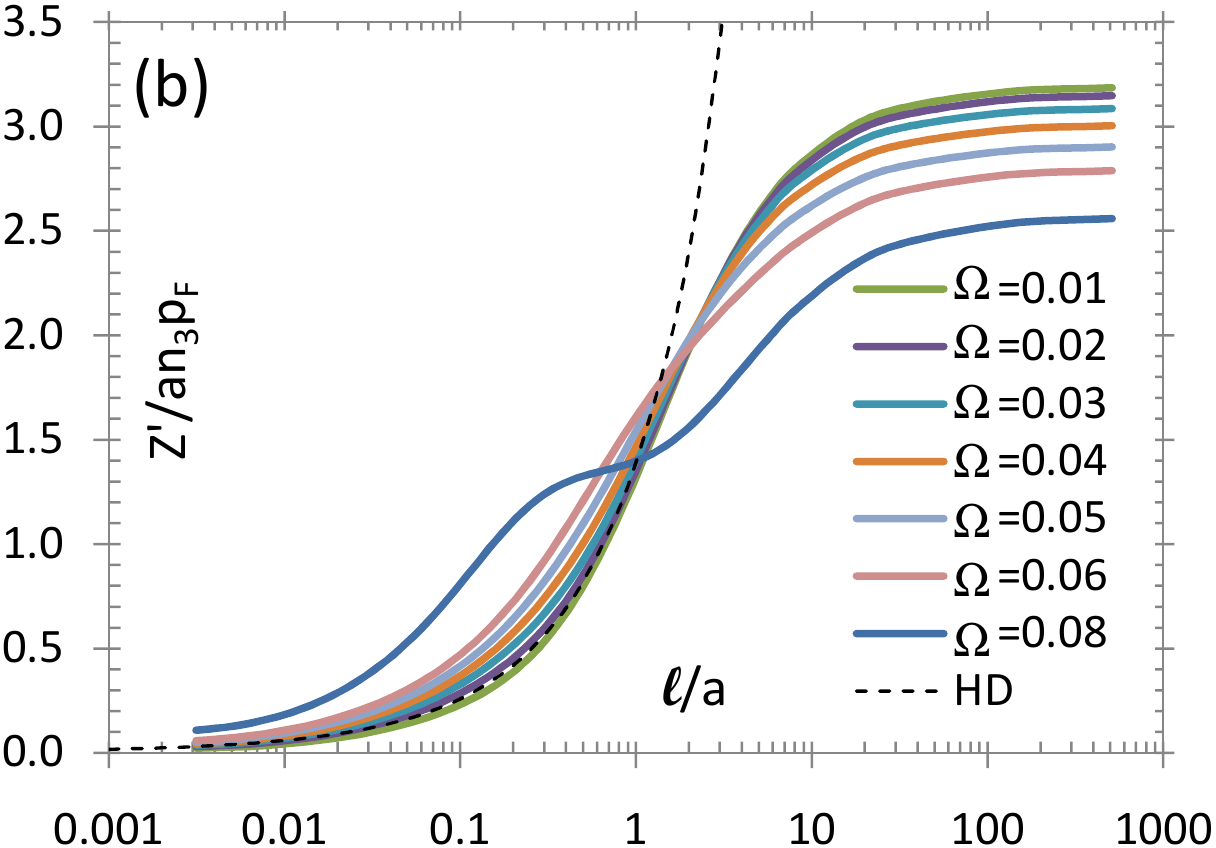}
\includegraphics[width=0.32\linewidth]{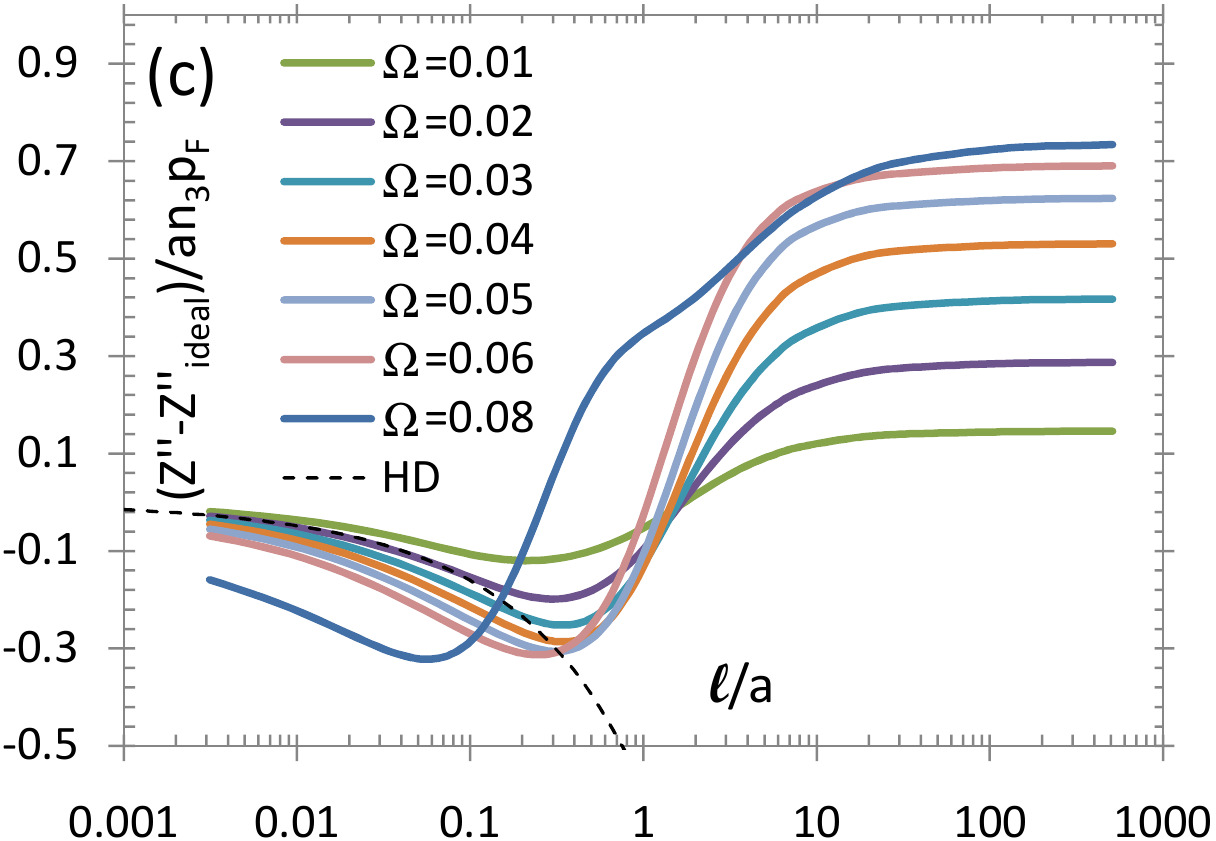}
} \caption{The effect of the frequency parameter $\Omega=a\omega/v_{\rm F}$. In this and the following figures \ref{fig-Zc}-\ref{fig-F1}, panel (a) shows $Z'$ vs. $Z''$ with $\ell/a$ as a parameter in the curves. Panels (b) and (c) show $Z'$ and $Z''$ as a function of $\ell/a$, the mean-free-path scaled by the radius of the wire. The variation of $\Omega$ can be understood as change of concentration, or change of the vacuum frequency of the wire. The dissipative part $Z'$ has a constant part independent of $\Omega$, but the reactive part $Z''$ vanishes for $\Omega\rightarrow 0$. For other parameters, the basic set is used: $b=10 a$, $S=0$, $F_0=-0.33$, $F_1=0.266$, and diffusive container walls. The curve for the largest value $\Omega=0.08$ differs essentially from the others because it is close to the lowest second-sound resonance, see Fig.\ \ref{fig-Zc}  for more details.  Here and in Figs.\ \ref{fig-Zc}-\ref{fig-F1} the dashed black curve gives the hydrodynamic (HD) Stokes solution (\ref{norm-sup-force}) for basic-set values of $\Omega$ and $F_1$.\label{fig-Z}}
\end{figure}

\begin{figure}[!tb]
\centerline{
\includegraphics[width=0.32\linewidth]{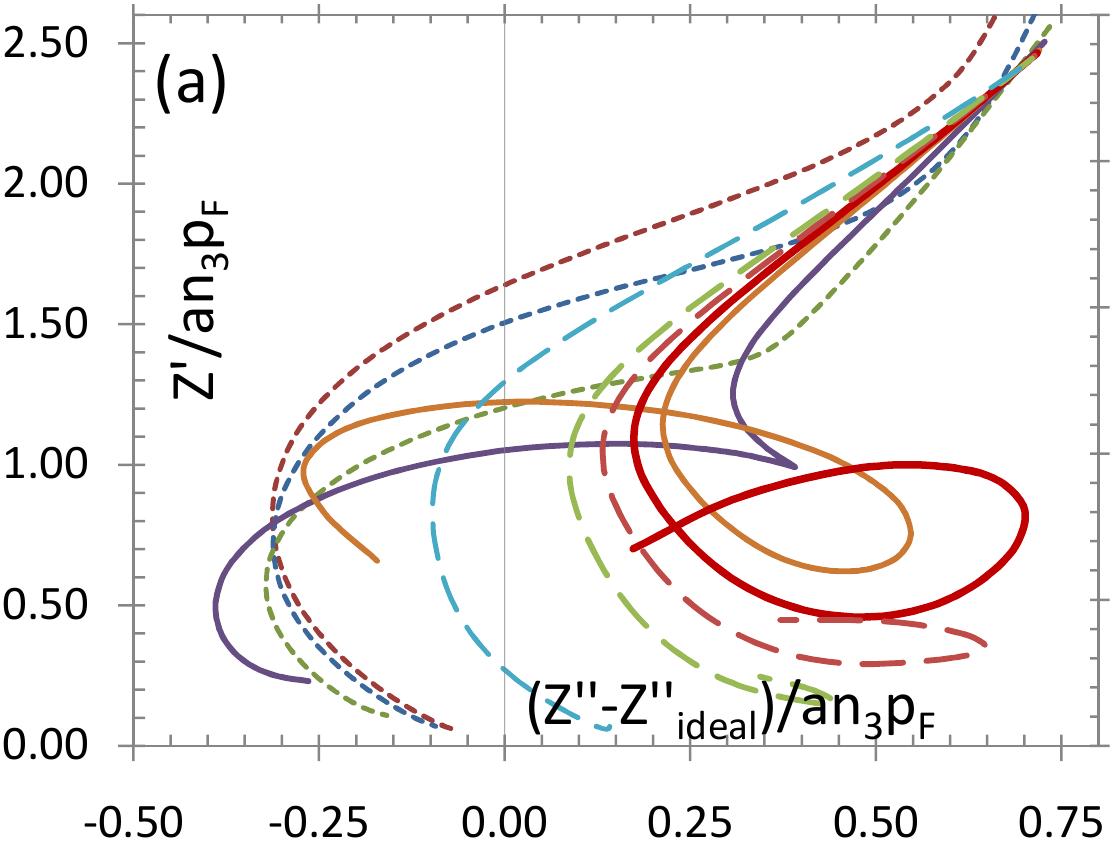}
\includegraphics[width=0.32\linewidth]{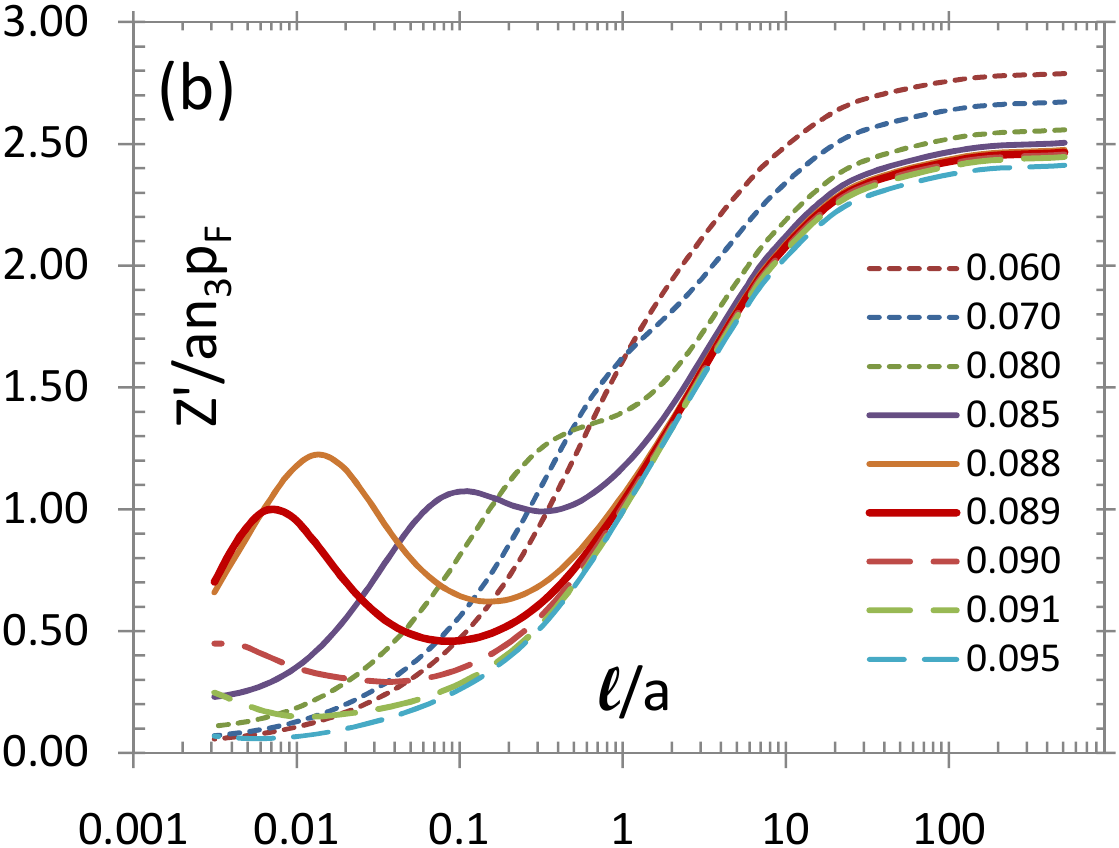}
\includegraphics[width=0.32\linewidth]{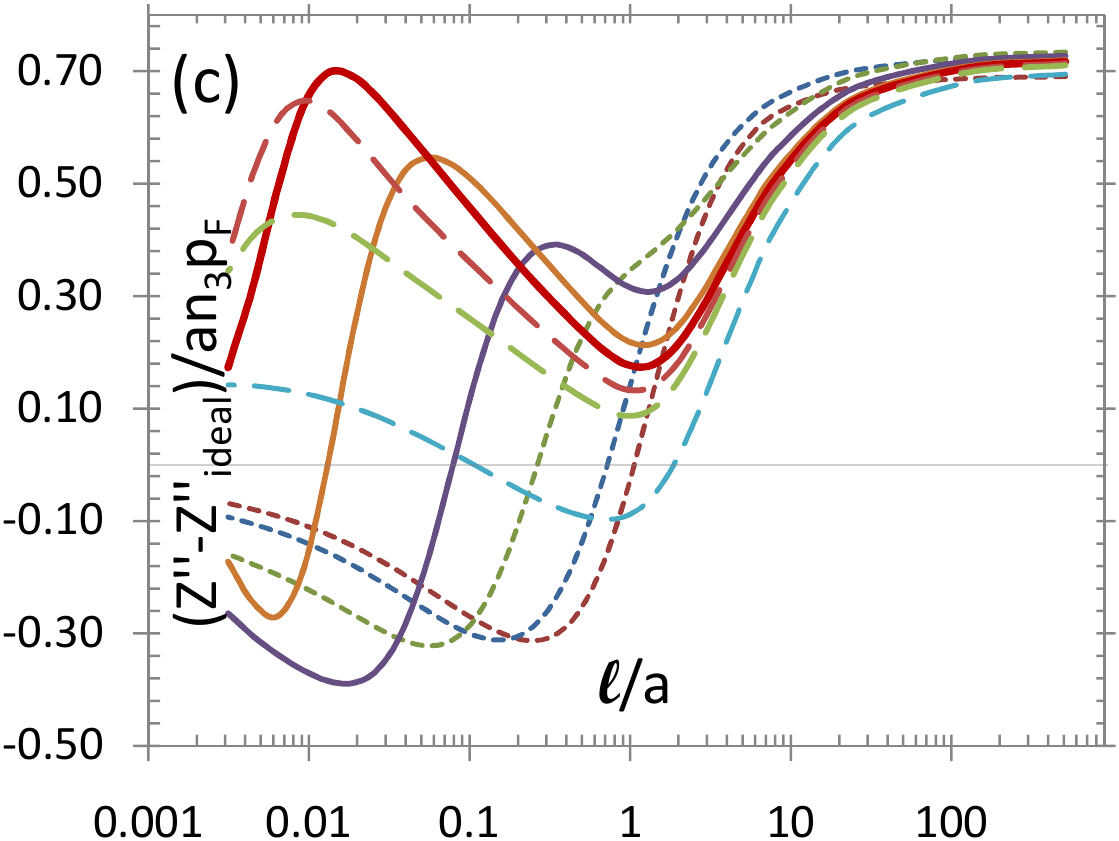}
} \caption{
Near the expected second sound resonance $\Omega_c=0.0890$ we
observe peculiar behavior in both dissipative and reactive parts of the
impedance. The three
short-dashed
lines with $\Omega=$ 0.06, 0.07, and 0.08 show the usual behavior below the
resonance. The three solid lines, $\Omega=$ 0.085, 0.088, and 0.089 are
close
to the resonance and make loops in the $Z'-Z''$ plane as seen in panel (a).
The long-dashed curves, $\Omega=$ 0.090, 0.091, and 0.095 show the behavior
above the resonance; it appears that these curves would form loops as well,
if we could extend the calculations to smaller $\ell$.
 For the resonance curve, $\Omega_c=0.089$, dissipative
part has a maximum at around $\ell=0.007 a$ and a minimum at $\ell\sim0.08
a$,
while the reactive part has a maximum at $\ell\sim0.015a$ and a minimum at
$\ell\sim 1.2a$. These extremal points move towards smaller $\ell$ with
increasing $\Omega$.
\label{fig-Zc}}
\end{figure}

The dependence of $Z$ on $\Omega=a\omega/v_{\rm F}$ is shown in Fig. \ref{fig-Z}. In the chosen dimensionless variables, the dependence on frequency $\omega$   appears through this parameter only. The reactive part $Z''$ vanishes for vanishing frequency. For the smallest values of $\Omega$ the dependence seems nearly linear, but the hydrodynamic limit is more complicated as discussed in Sec.\ \ref{sec-hd}. The dissipative part $Z'$ has weaker dependence on frequency and reduces to a finite value at $\Omega\rightarrow 0$, except in the limit $\ell\rightarrow0$.  At larger $\Omega$ nonlinearity appears in $Z''$ at all $\ell/a$. The last curve at $\Omega=0.08$ differs essentially from the others. This behavior is caused by $b\omega/v_{\rm F}$ approaching unity, where the lowest second-sound resonance appears in the container. More precisely, the lowest pole of Eq.\ (\ref{e.zserosre}) gives the resonance at $\Omega_c=0.0890$. The impedance in the neighborhood of the resonance is shown in Fig.\ \ref{fig-Zc}. Resonance features that resemble our results have been observed in experiments using  quartz tuning forks, that work at  considerably higher  frequencies (33 kHz) [\onlinecite{Pentti09}].

The resonances in Fig.\ \ref{fig-Zc} appear strongest at small $\ell$ while they are damped at larger $\ell$.
The behavior of the ballistic limit point is shown in Fig.\ \ref{f.balom}. Instead of sharp resonances one sees oscillatory behavior. Thus we see complete change-over from the second sound resonances in the hydrodynamic regime to quasiparticle interferences in the ballistic regime.

\begin{figure}[!tb]
\centerline{
\includegraphics[width=0.3\linewidth]{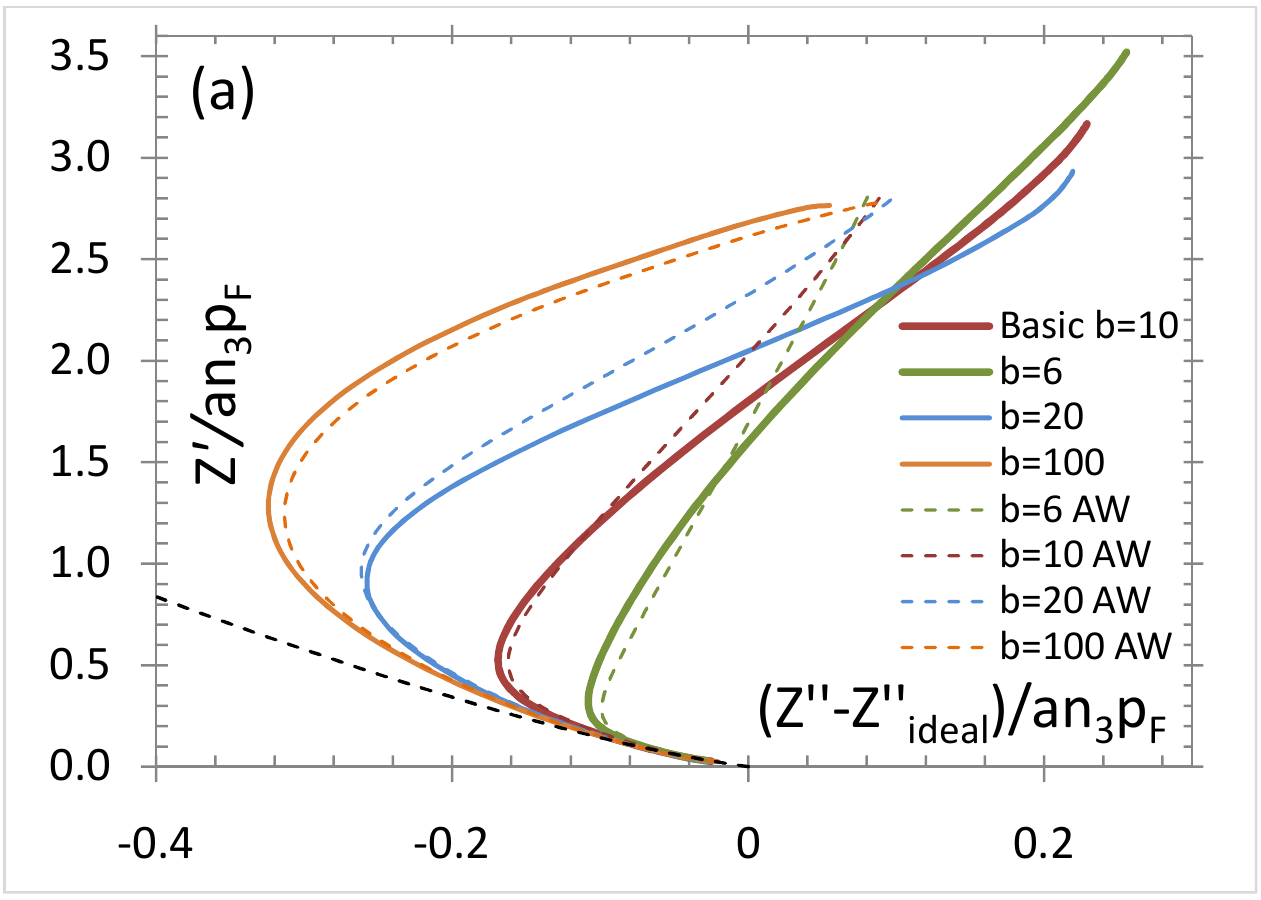}
\includegraphics[width=0.3\linewidth]{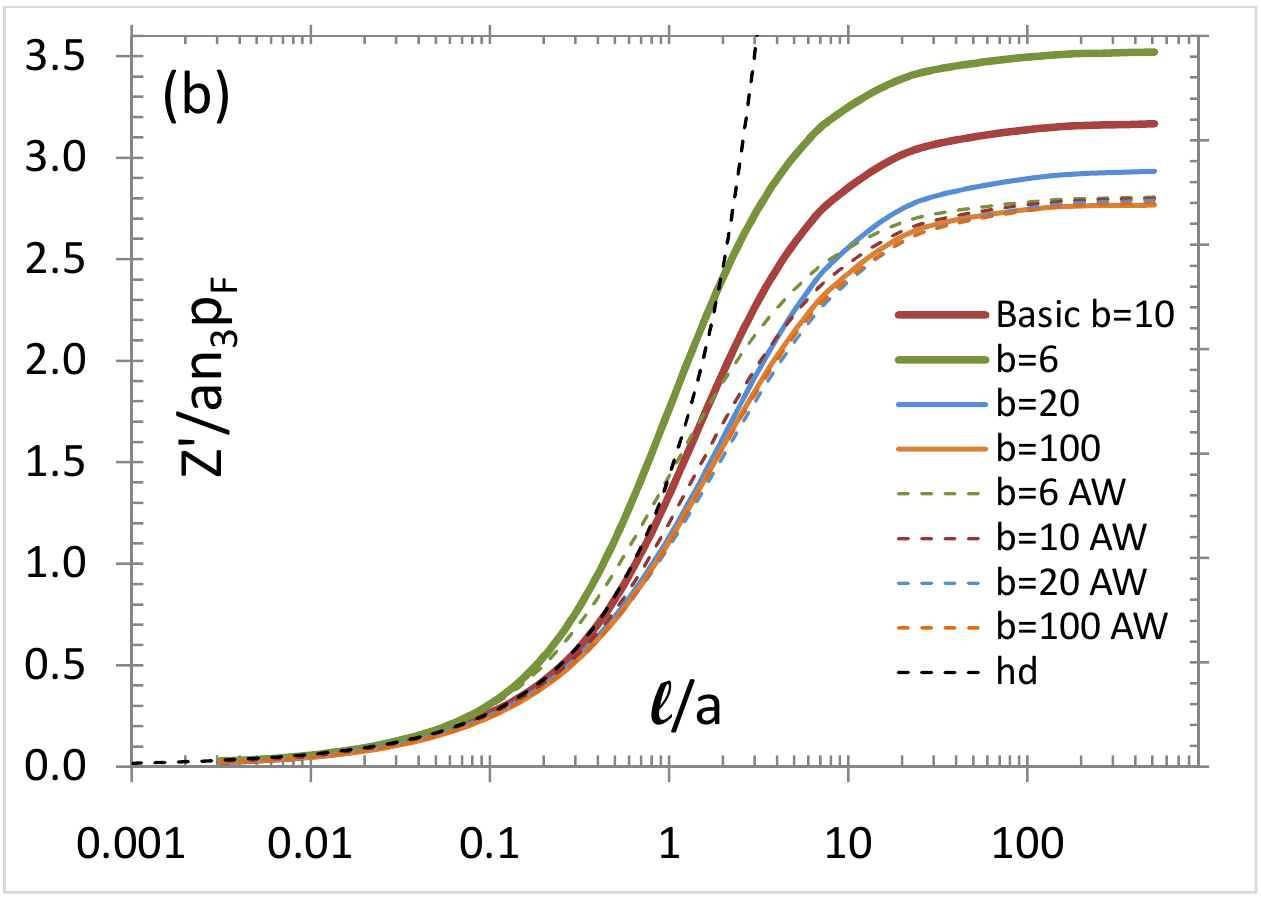}
\includegraphics[width=0.3\linewidth]{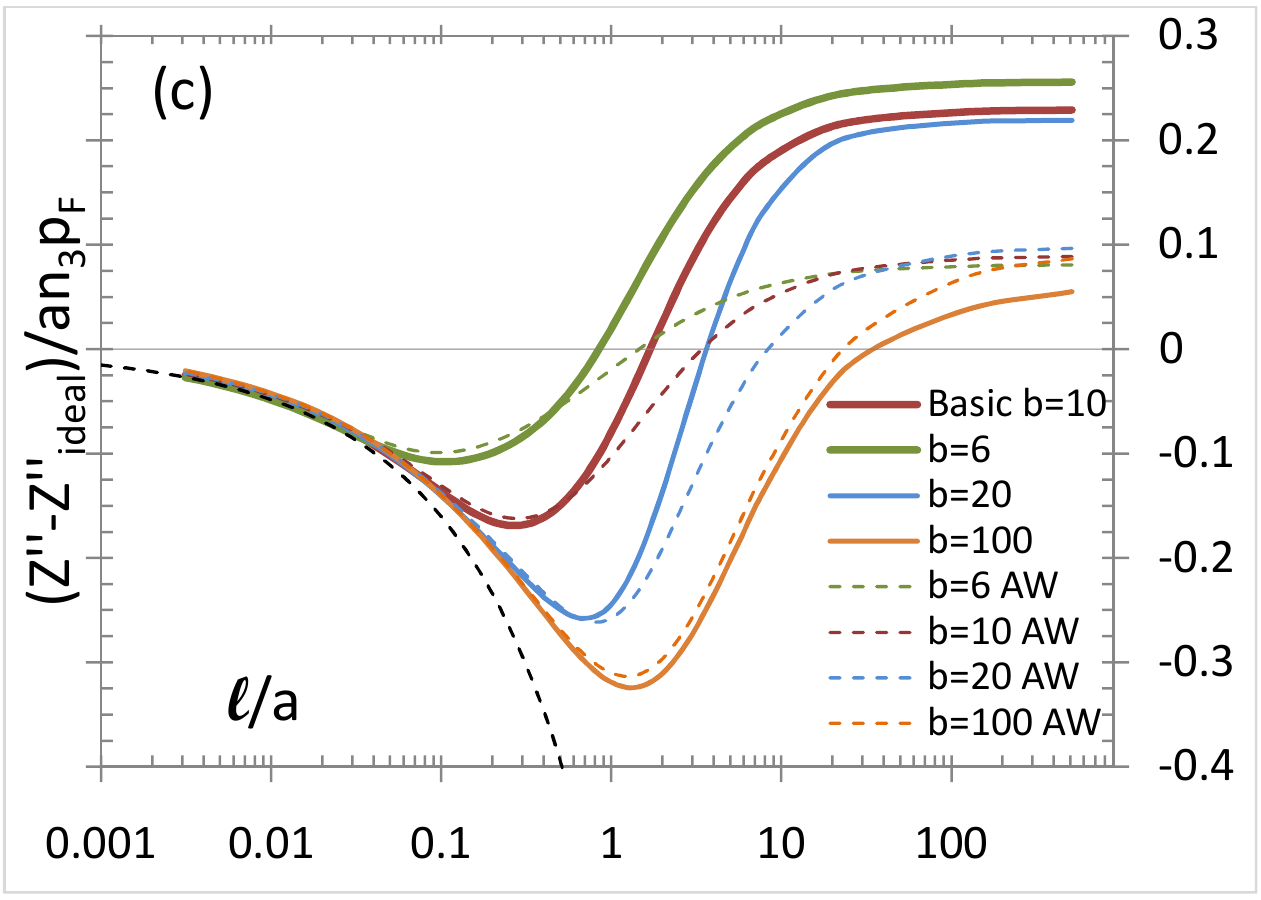}
} \caption{The effect of the container radius $b$. The solid lines correspond to diffuse container walls, while the dashed lines correspond to absorbing walls. For small $\ell$ the container radius has only small effect, and the curves merge. At larger $\ell$, the maximum shift of frequency is larger for larger container, while in the ballistic limit the maximum dissipation is larger for smaller container. The largest radius corresponds to effectively unlimited fluid. We see that although $Z''>0$ for $b=100 a$ in the ballistic limit, it is quite small in comparison to the smaller containers. In the case of absorbing walls, the curves end to the same point in the ballistic limit, close to the end point of the large container curve with diffuse walls.  \label{fig-b}}
\end{figure}

The effect of confinement is studied in Fig.\ \ref{fig-b} in terms of the radius $b$ of the cylindrical container. For diffusive container walls $b$ has a strong effect on the results: the dissipation and the minimum frequency are larger for smaller containers. If we use absorbing walls for the container, the container size does not have such a drastic effect. It is understood that the reflecting walls allow for second-sound resonances, which occur at $b\Omega/v_{\rm F}\sim 1$, and have a large effect at higher frequencies or at larger container sizes.

\begin{figure}[!tb]
\centerline{
\includegraphics[width=0.3\linewidth]{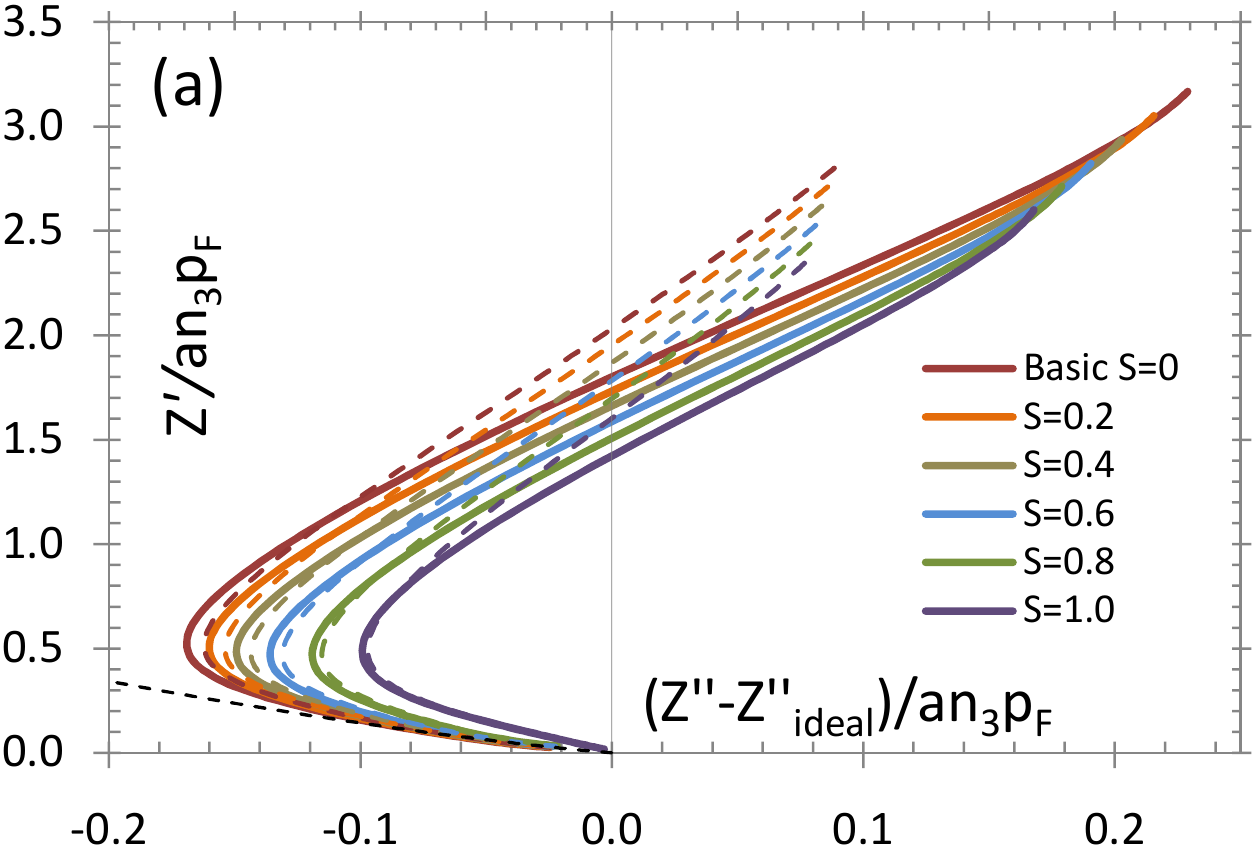}
\includegraphics[width=0.3\linewidth]{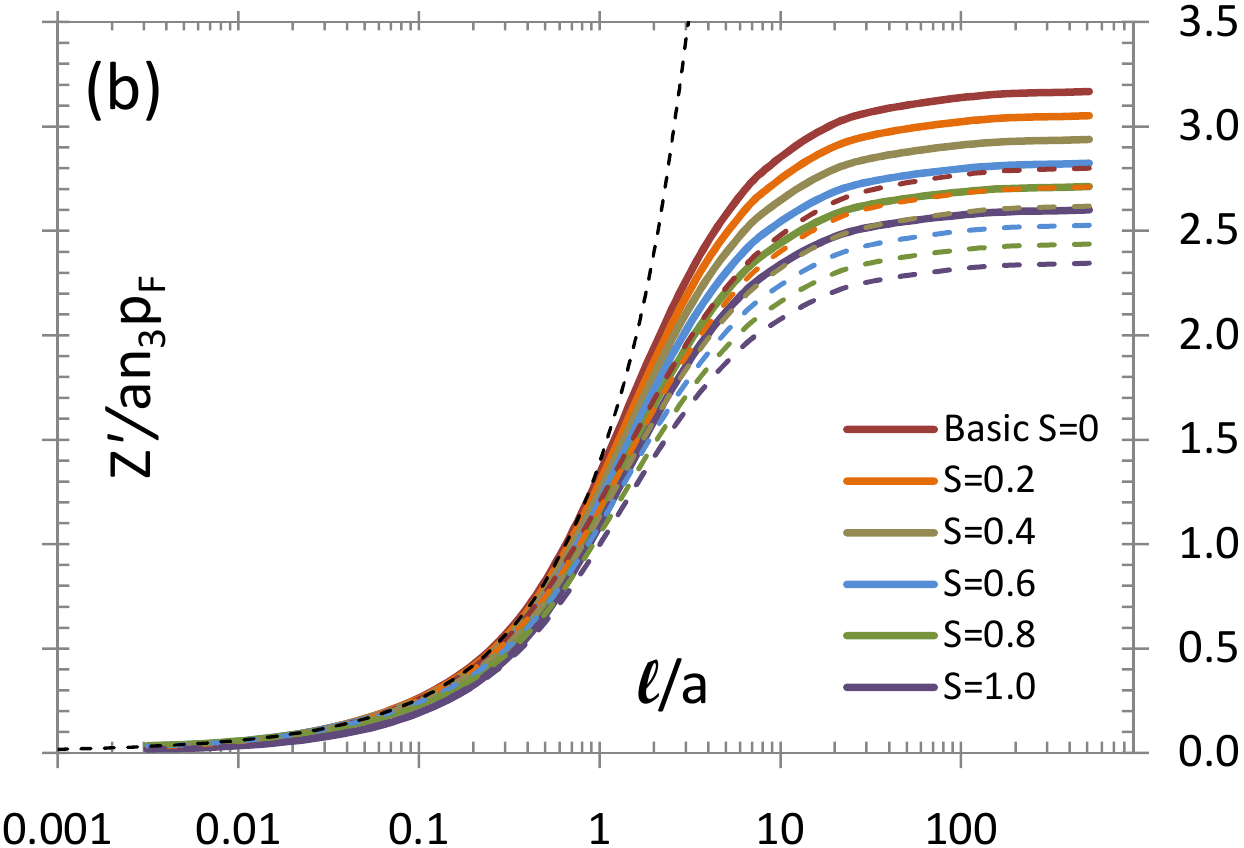}
\includegraphics[width=0.3\linewidth]{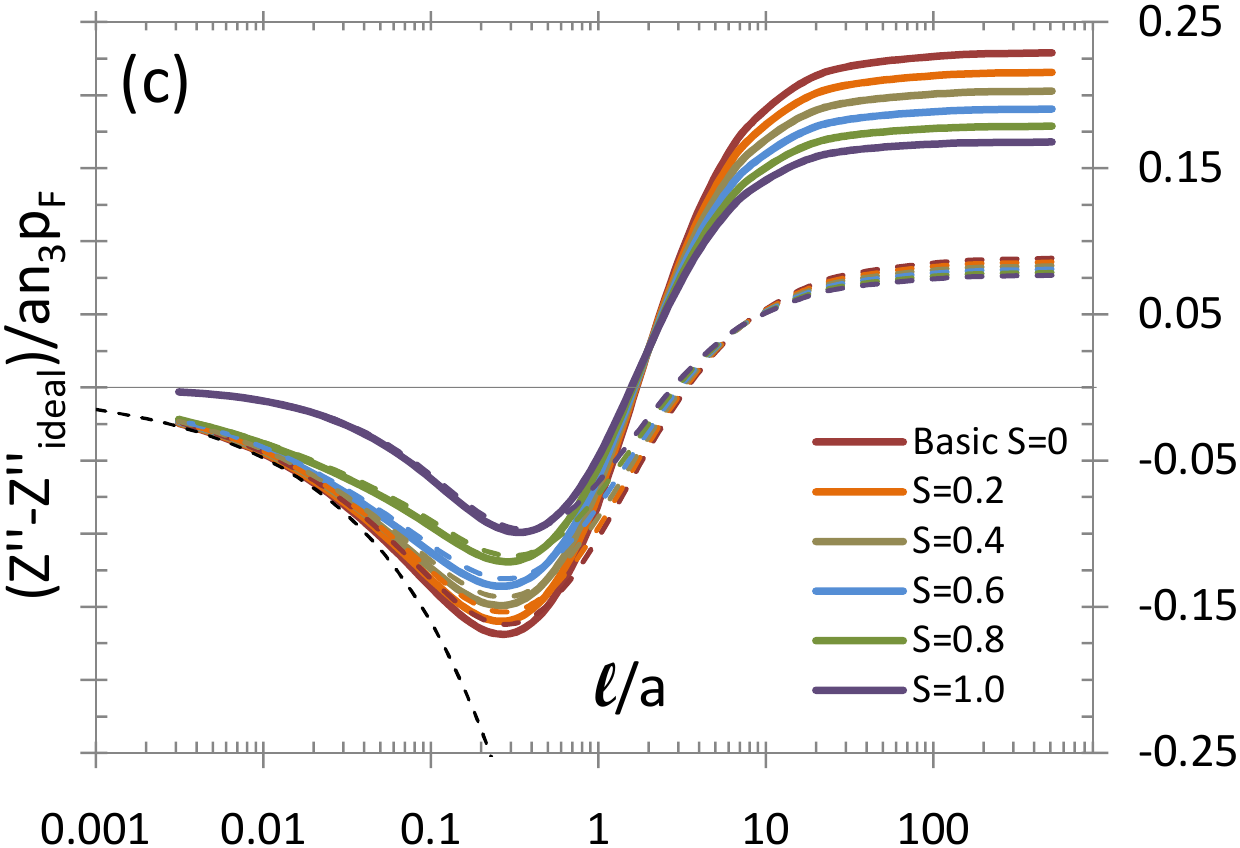}
} \caption{The effect of the specularity $S$ of the boundary condition at the wire surface. For small $\ell$ the effect of $S$ is hardly visible, except for the purely specular curve $S=1$ which stands out in the hydrodynamic limit for $Z''$. At larger $\ell$ the dependence on $S$ is uniform in both $Z'$ and $Z''$. Both the maximum frequency shift and the maximum dissipation are larger for a diffusive wire.
\label{fig-S}}
\end{figure}

The dependence on the specularity parameter $S$ of the wire surface is shown in Fig.\ \ref{fig-S}. Increasing $S$ means increasing slippage on the wire surface, and leads to less fluid moving with the wire. Then the dissipation and the change of resonant frequency are smaller than for fully diffuse wire. We emphasize that the parameter $S$ is assumed to be independent of $\ell$. We see that the dependence on $S$ is nearly linear  except in the hydrodynamic region, where the fully specular case $S=1$ stands out from all other values $S<1$. For absorbing container walls the effect on the resonant frequency in the ballistic limit is small.

\begin{figure}[!tb]
\centerline{
\includegraphics[width=0.3\linewidth]{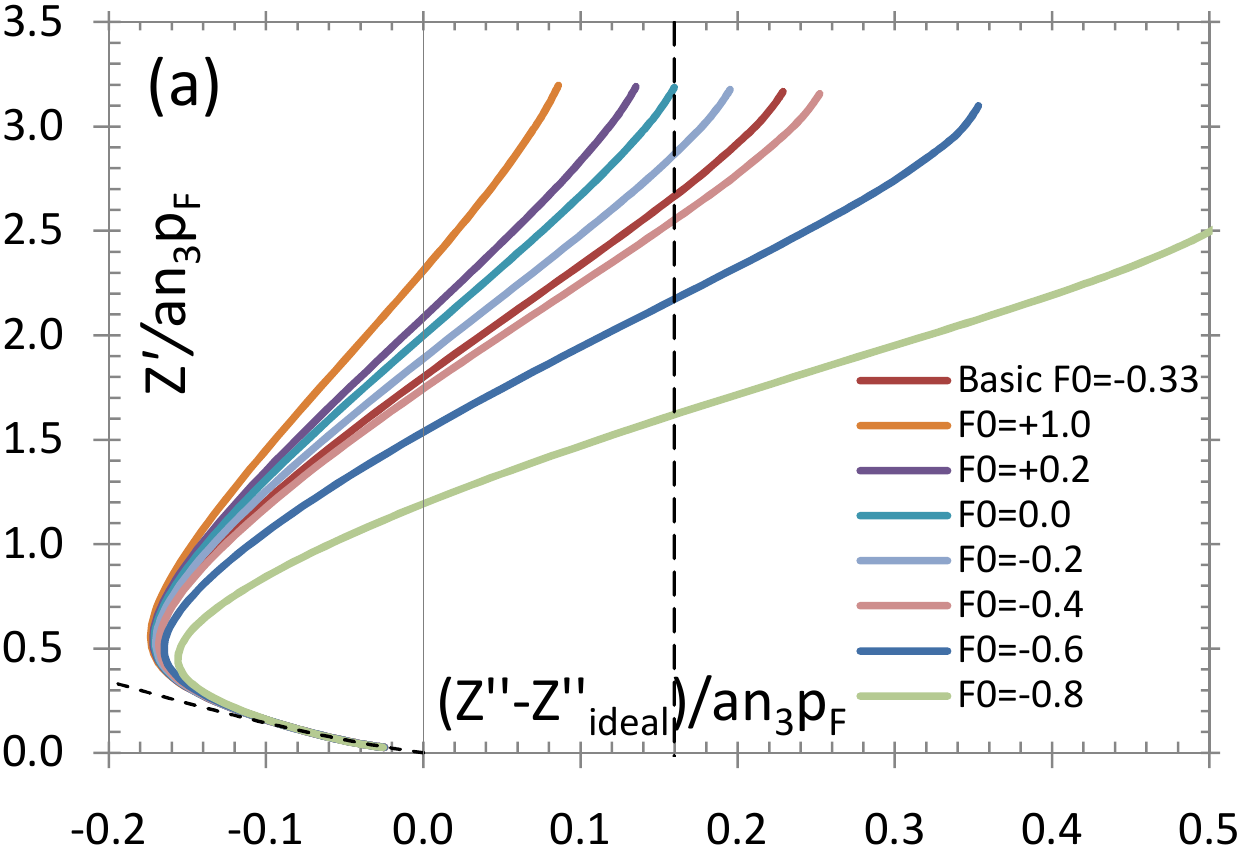}
\includegraphics[width=0.3\linewidth]{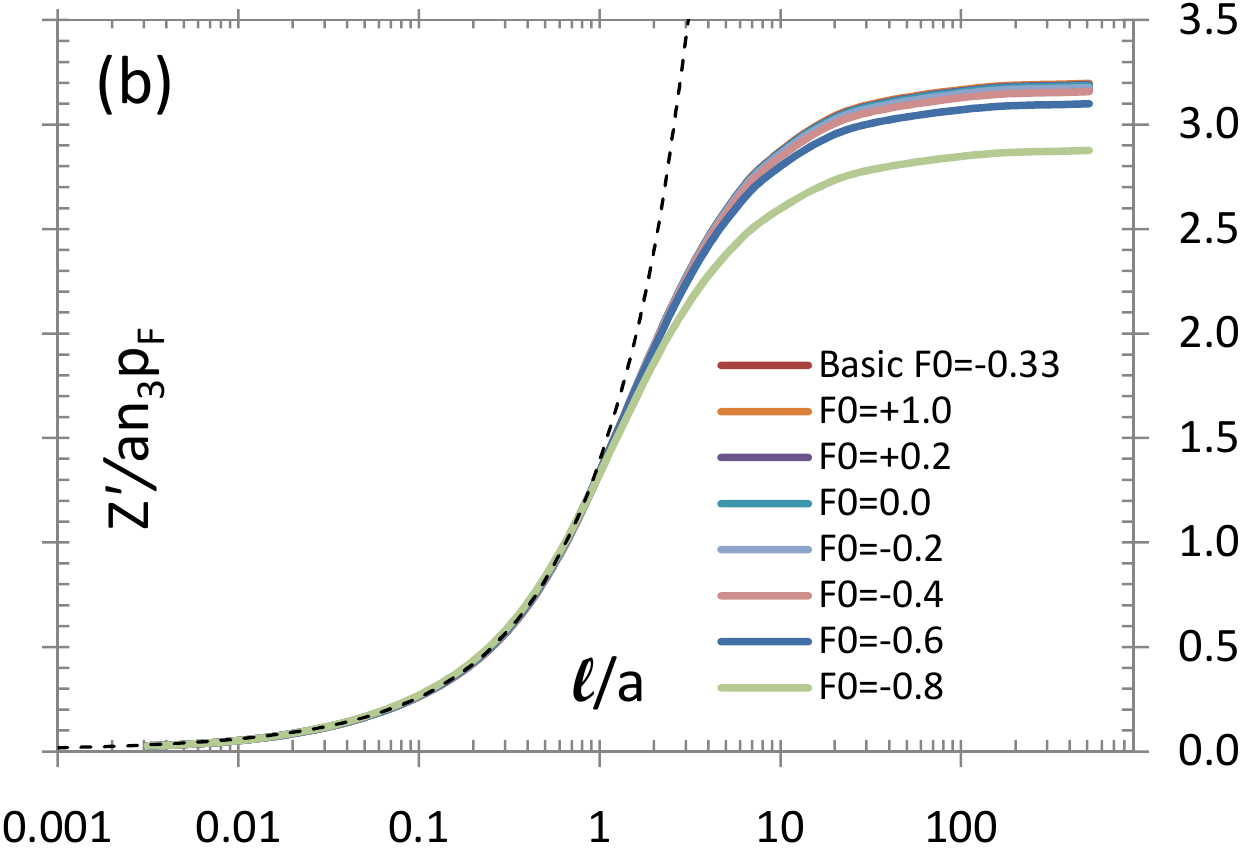}
\includegraphics[width=0.3\linewidth]{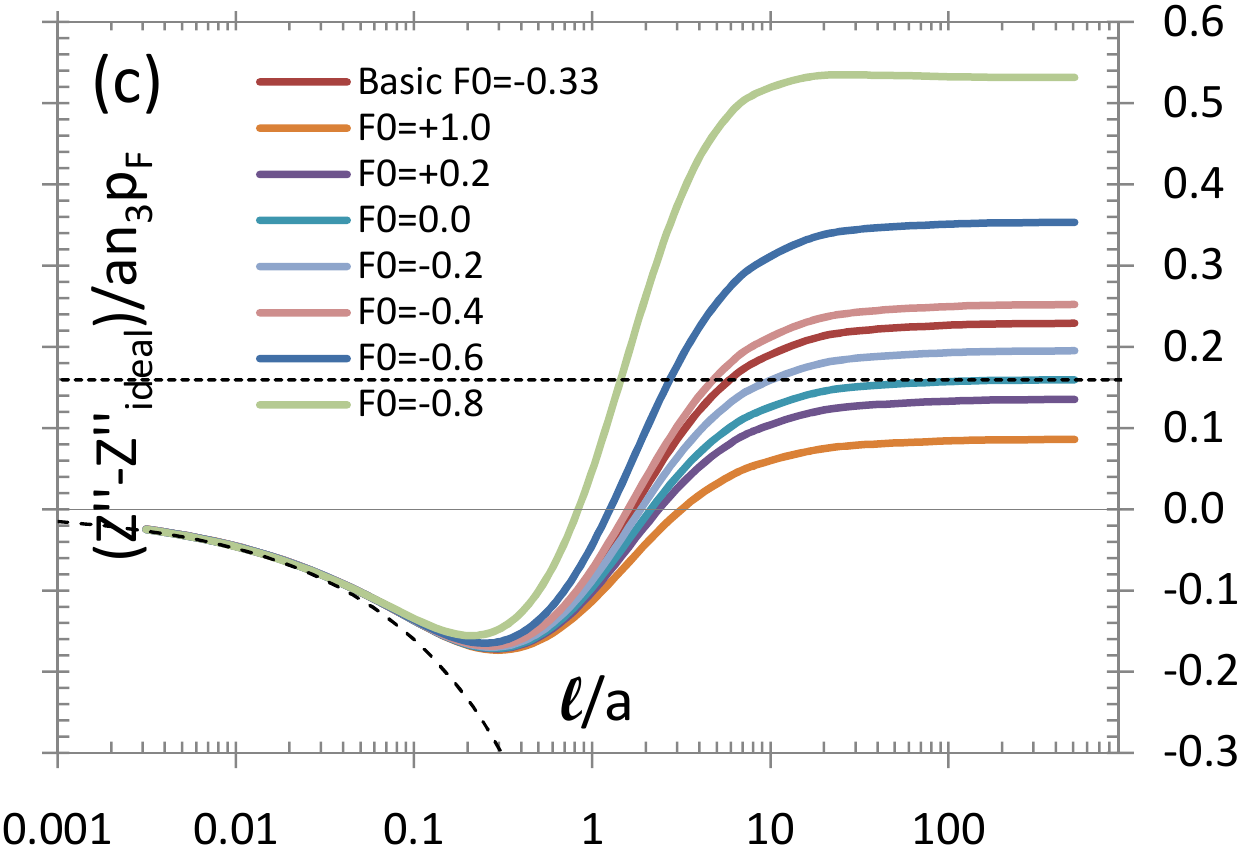}
} \caption{The effect of  $F_0$. In the $Z'$ vs. $Z''$ curve, $F_0$ controls the end point and the slope of the ballistic branch of the curve, while its effect is small in the hydrodynamic region. From panel (b) we see that the effect on the dissipation is small, apart from the smallest value shown here, $F_0=-0.8$ in the ballistic regime. The effect on the frequency is larger, as discussed in Ref.\ \onlinecite{VTletter}. The vertical (a) and horizontal (c) dashed lines mark the ballistic limit value of $Z''$ in the case of $F_0=0$. \label{fig-F0}}
\end{figure}

The contribution of the quasiparticle interactions to $Z$ is called the Landau force \cite{VTletter}. 
The dependence on the interaction parameter $F_0$ is studied in Fig.\ \ref{fig-F0}. The effect of $F_0$ appears via the change of density. In the hydrodynamic regime the density stays almost constant and therefore the effect of $F_0$ is small. In the ballistic regime density changes are essential and $F_0$ has a large effect on $Z''$. For stability $F_0$ has to be larger than $-1$. The effect of $F_0$  on $Z''$ seems to be amplified for $F_0$ approaching the instability. 
For a low frequency such as in Fig.\ \ref{fig-F0}, the effect of $F_0$ on the real part $Z'$ is small. With increasing frequency the main effect turns from $Z''$ to  $Z'$, as can be seen in the ballitic-limit Fig.\ \ref{f.balom}. In the large frequency limit the main effect of the Landau force is the reduction of  $Z'$ (for  $F_0<0$).

\begin{figure}[!tb]
\includegraphics[width=0.3\linewidth]{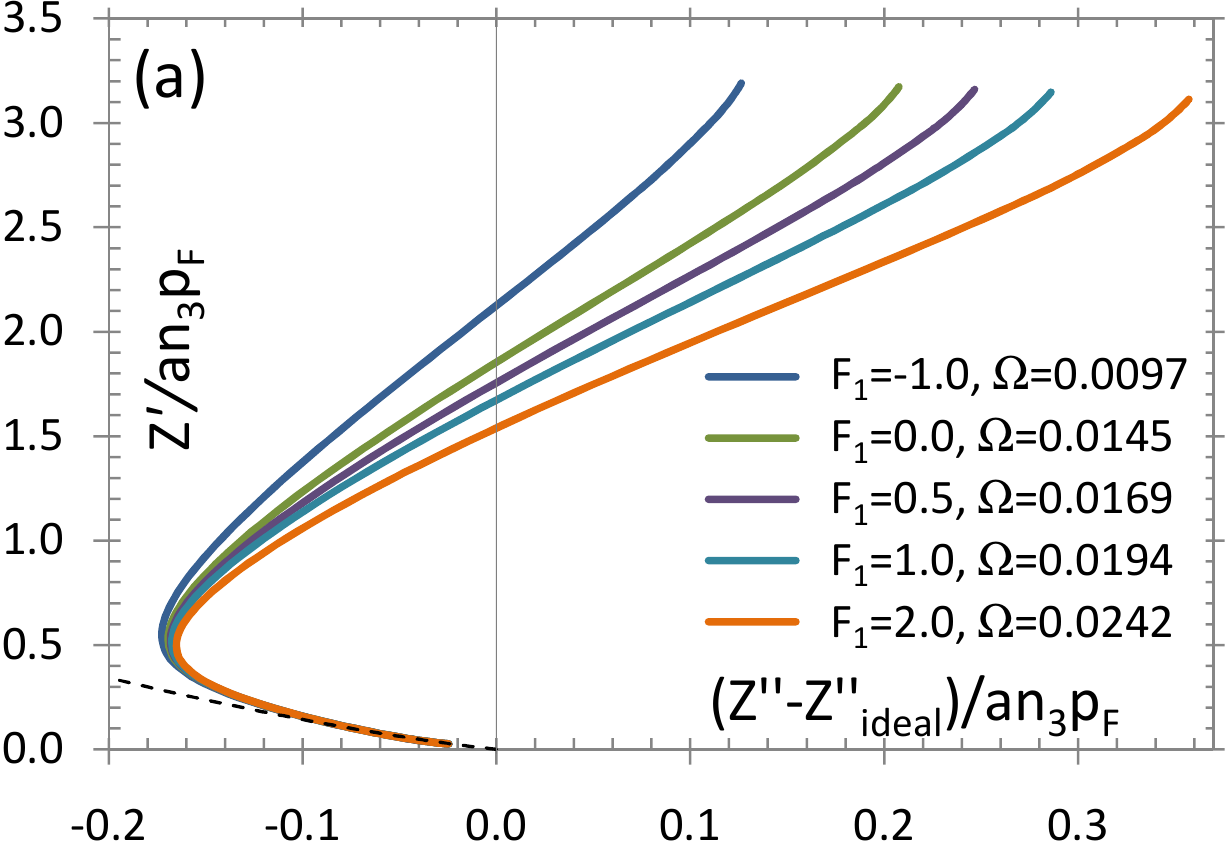}
\includegraphics[width=0.3\linewidth]{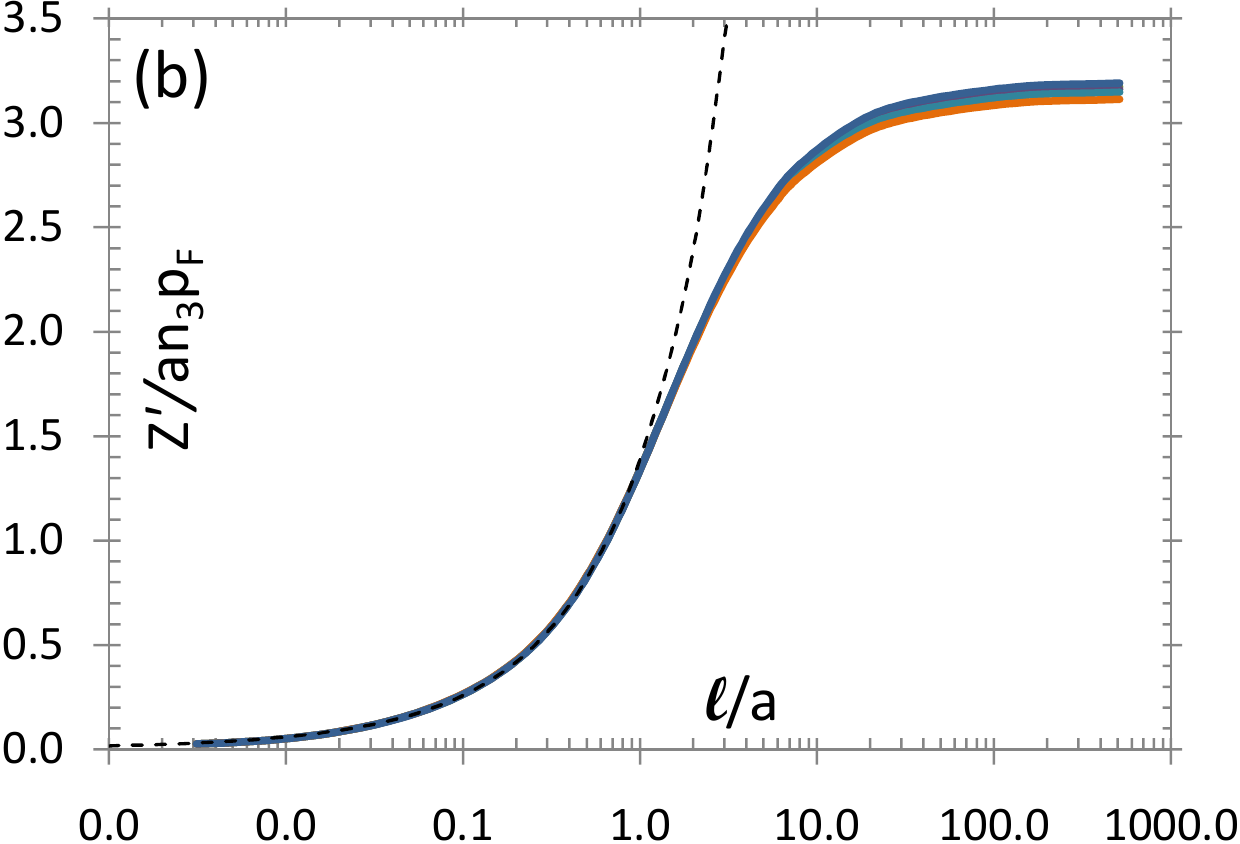}
\includegraphics[width=0.3\linewidth]{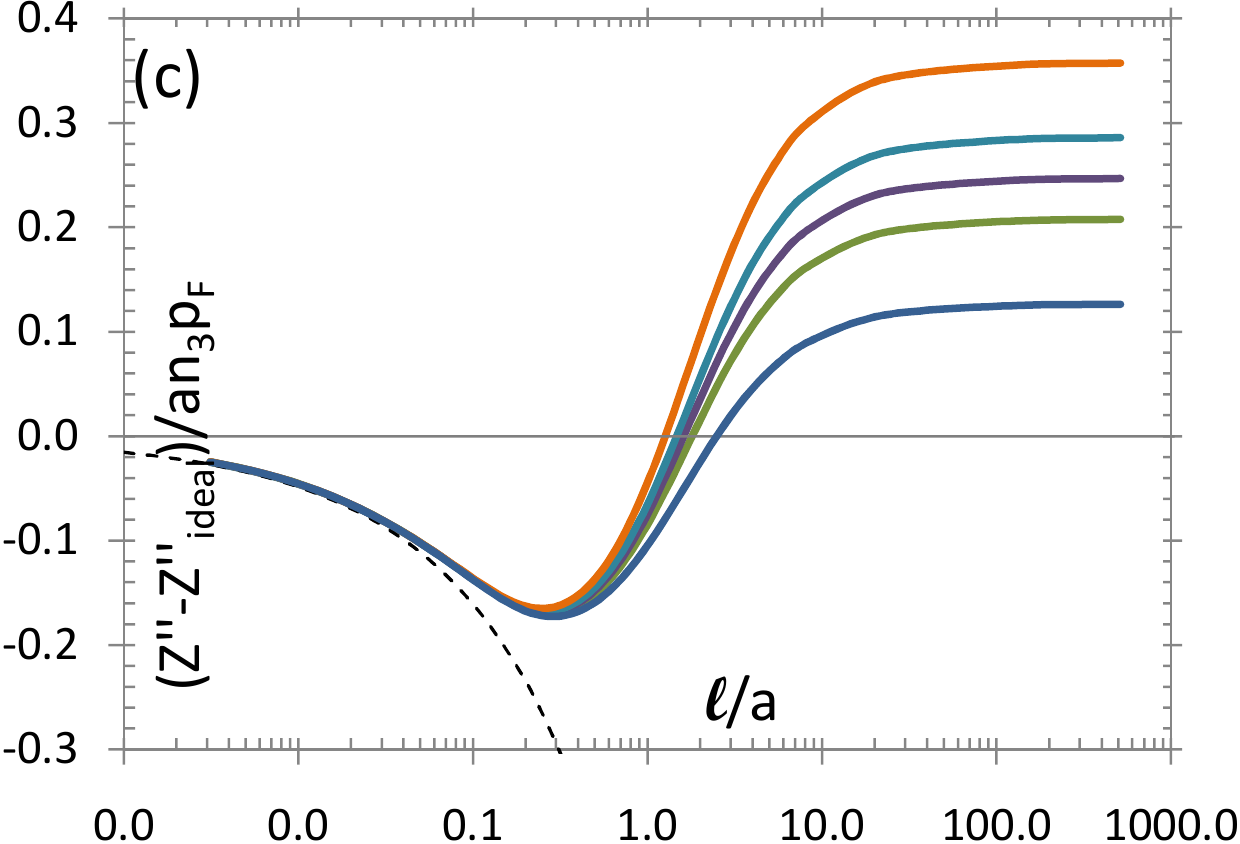}
\caption{The effect of $F_1$ at constant $\Omega/(1+F_1/3)$. The dependence is similar as on  $F_0$ (Fig.\ \ref{fig-F0}) when parameters $F_1$ and $F_0$ are changed in opposite directions. The curve labeled $F_1=-1$ would correspond to $m^*<m_H$.\label{fig-F1}}
\end{figure}

The effect of the second interactions parameter $F_1$ is shown in Fig.\ \ref{fig-F1}. For constant $\Omega/(1+F_1/3)$ its effect is very similar to that of $F_0$, the curves almost coincide. The difference is that increasing $F_1$ is compensated by a smaller decrease of $F_0$. The opposite tendency can be understood based on the qualitative explanation of the Landau force given in Ref.\ \onlinecite{VTletter}. Namely, the oscillating wire creates a beam of quasiparticles that affects the quasiparticles that are incident on the wire. The relevant interaction thus has $\hat{\bm p}\cdot\langle \hat{\bm p}'<0$ in Eq.\ (\ref{e.deltae2}) and thus  $F_0$ and $F_1$ appear with weights of opposite signs.

We notice that the dissipative part $Z'$  depends weakly on $F_0$, $F_1$, and $\Omega/(1+F_1/3)$ (except close to resonance), while the reactive part $Z''$ has a stronger dependence on the parameters. For $S$ the both parts are affected in a similar fashion, and for $b$ the dependence is rather complicated.

\begin{figure}[tbh]
\includegraphics[width=0.3\linewidth]{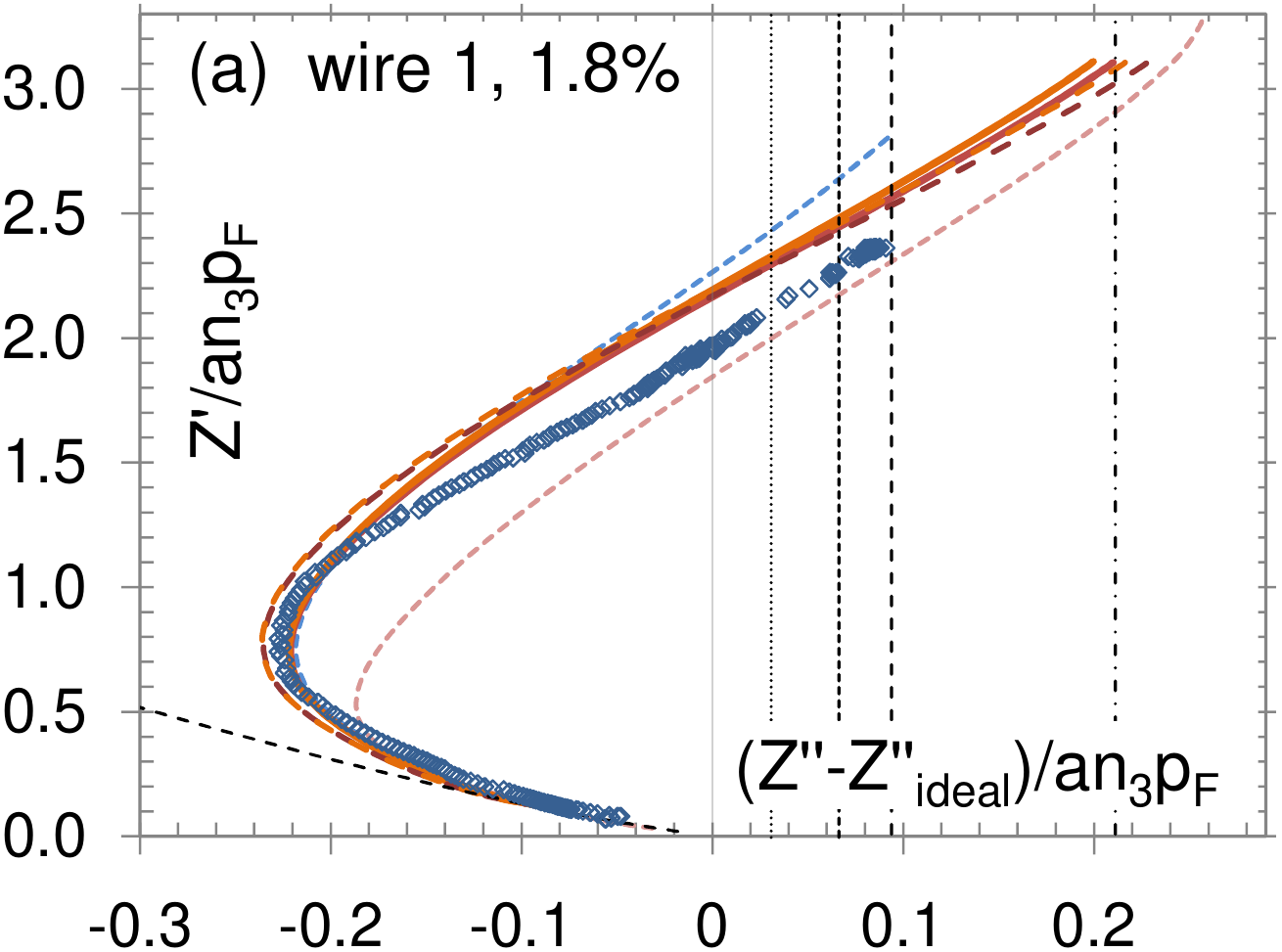}
\includegraphics[width=0.3\linewidth]{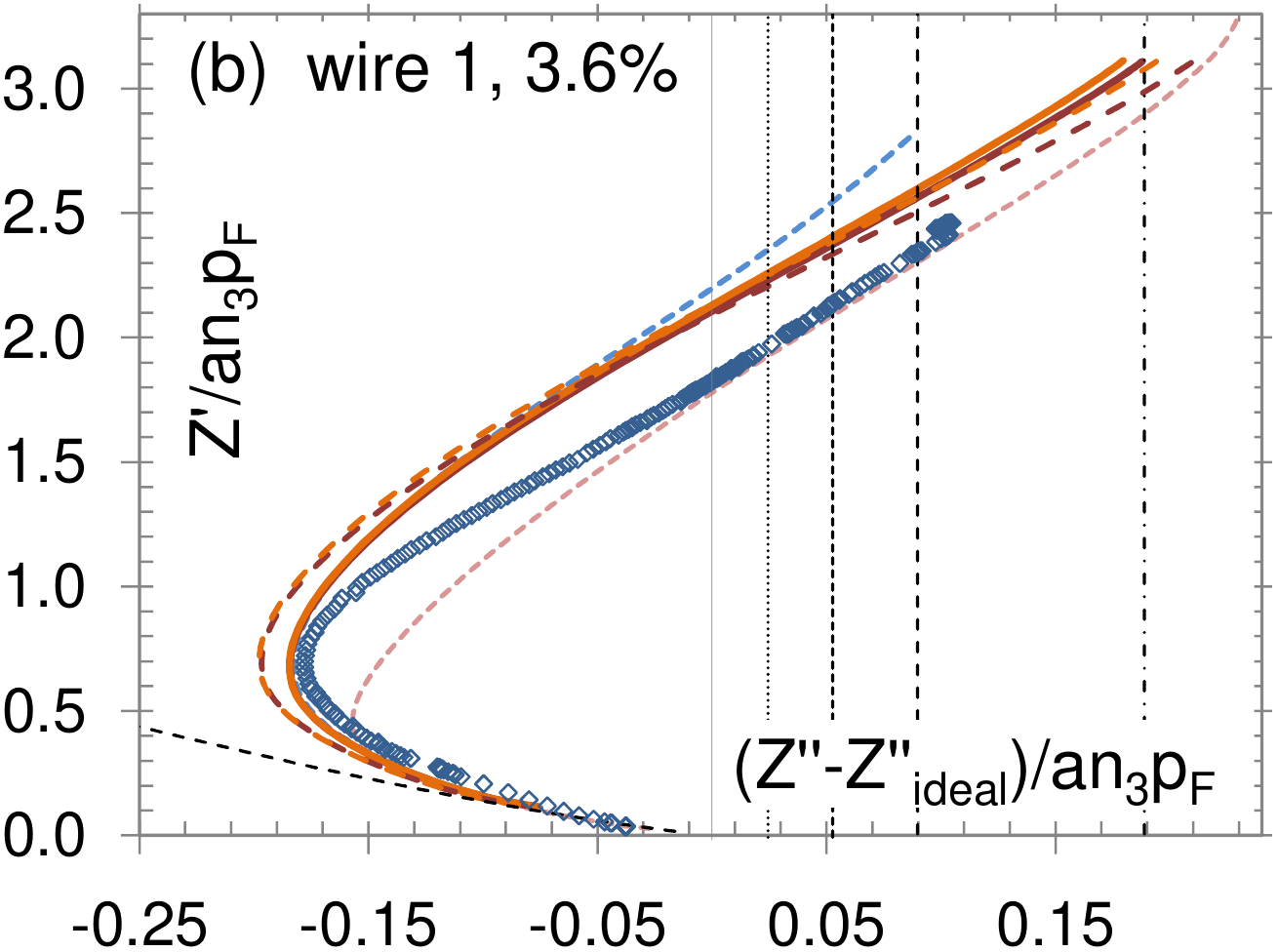}
\includegraphics[width=0.3\linewidth]{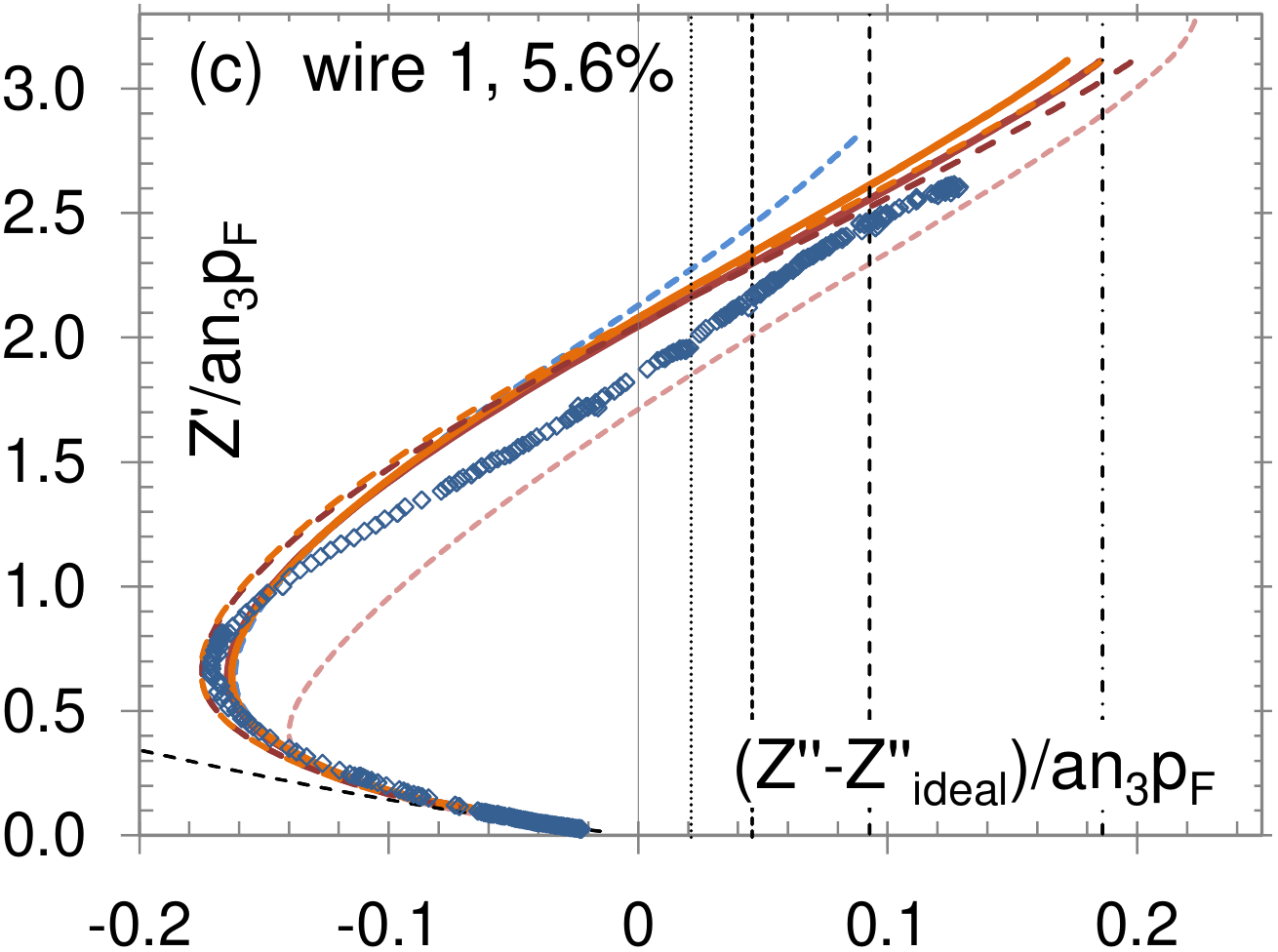}

\vspace{0.5cm}
\includegraphics[width=0.3\linewidth]{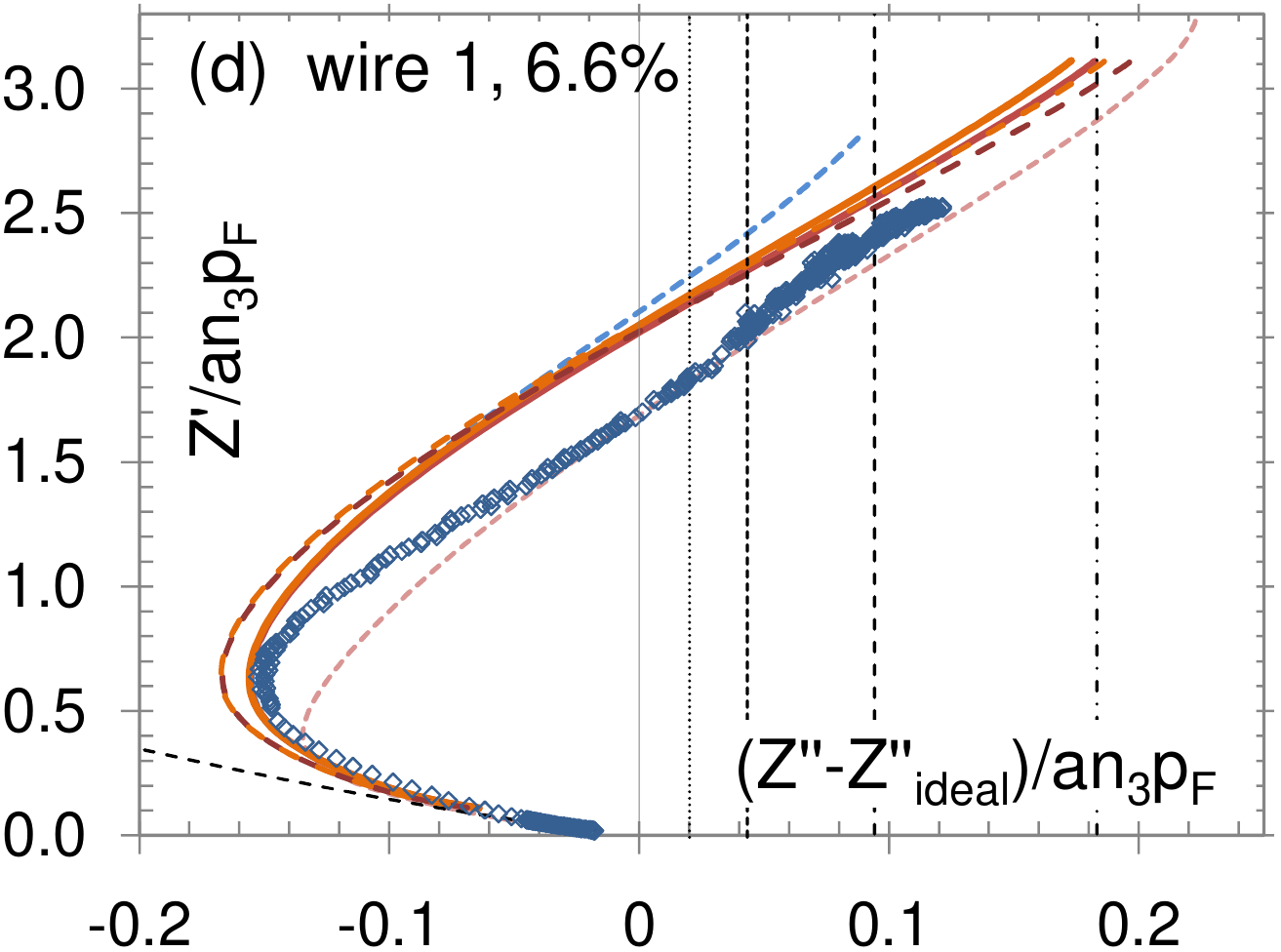}
\includegraphics[width=0.3\linewidth]{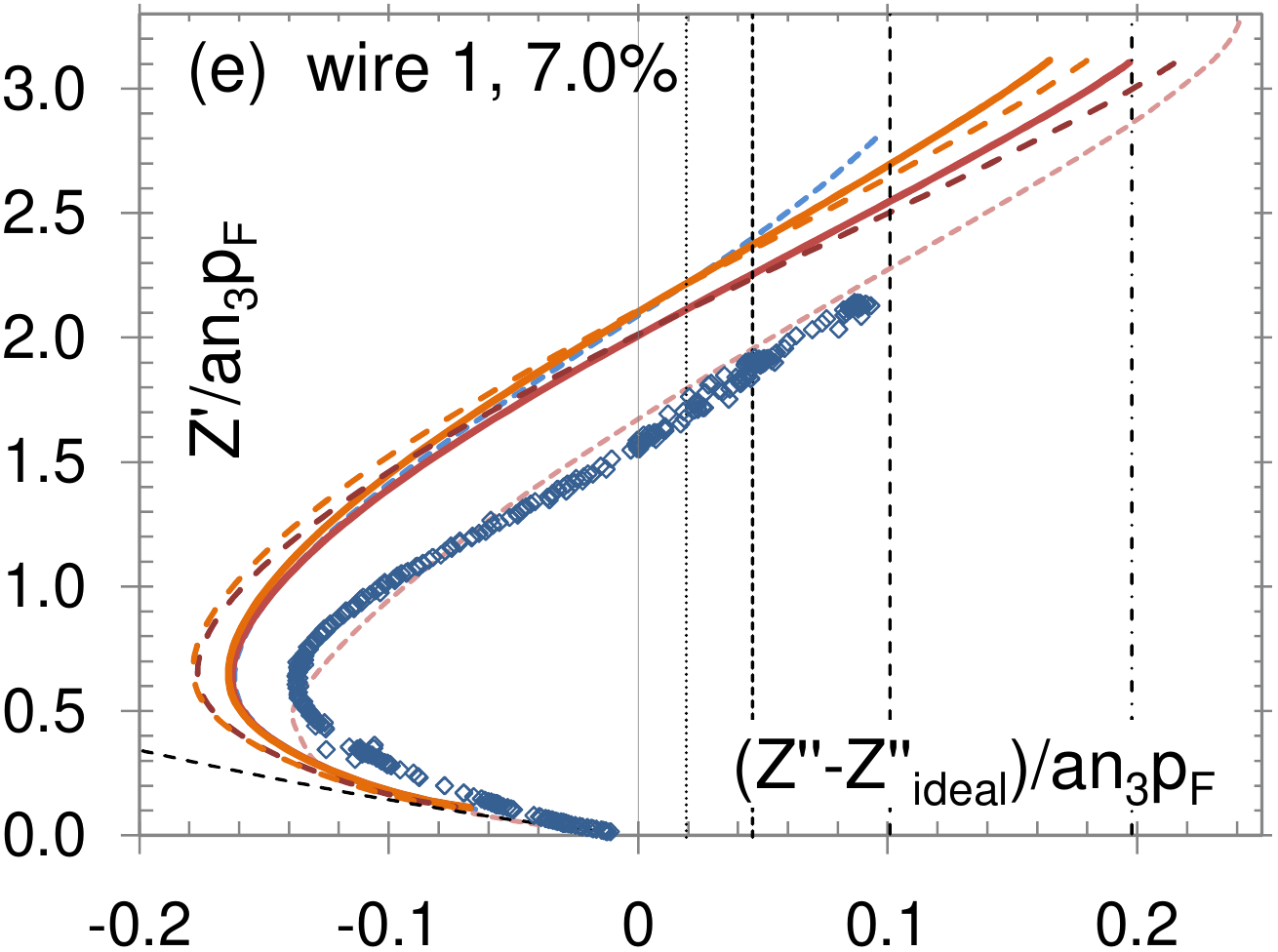}
\includegraphics[width=0.3\linewidth]{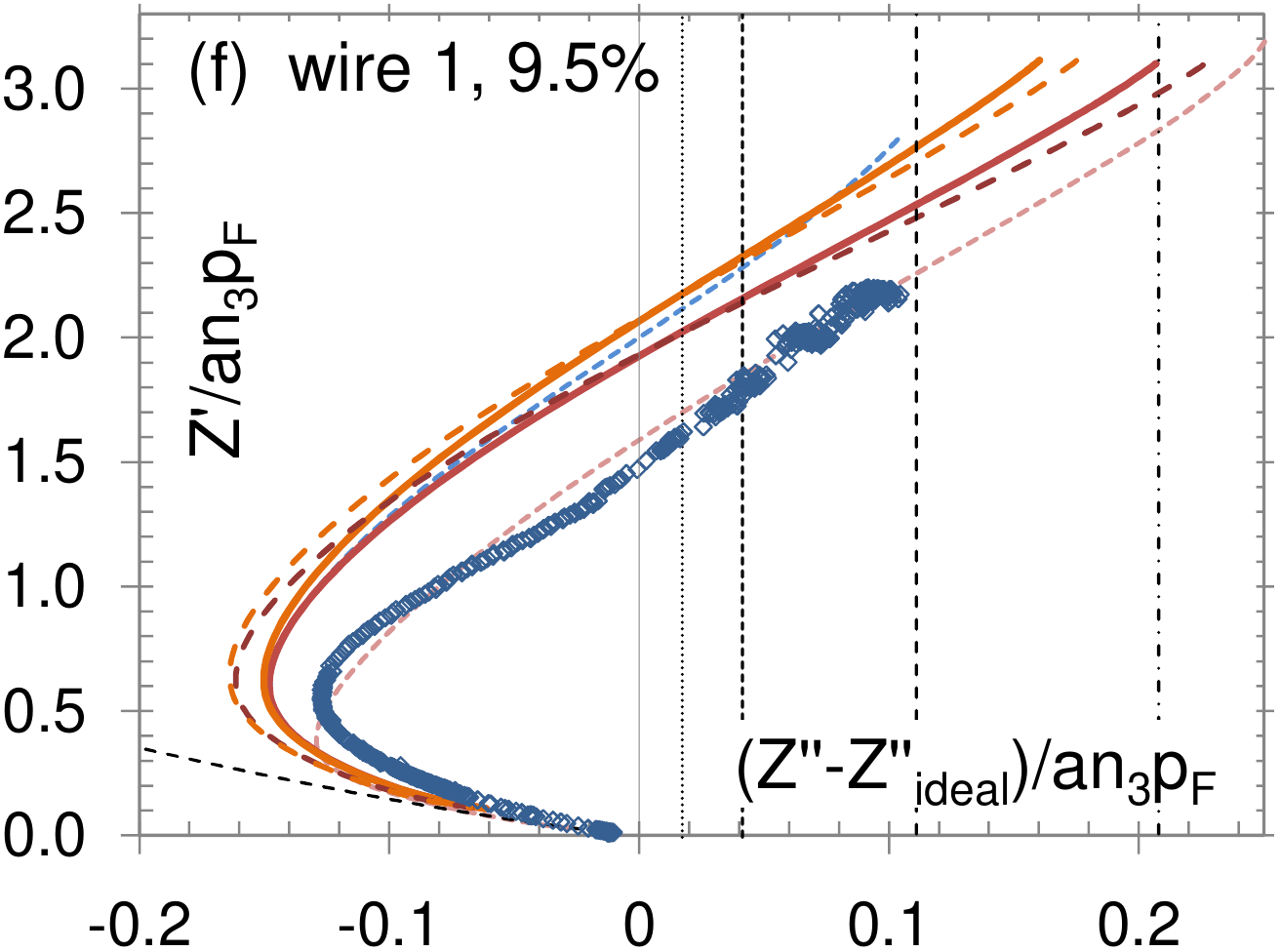}

\vspace{0.5cm}
\includegraphics[width=0.3\linewidth]{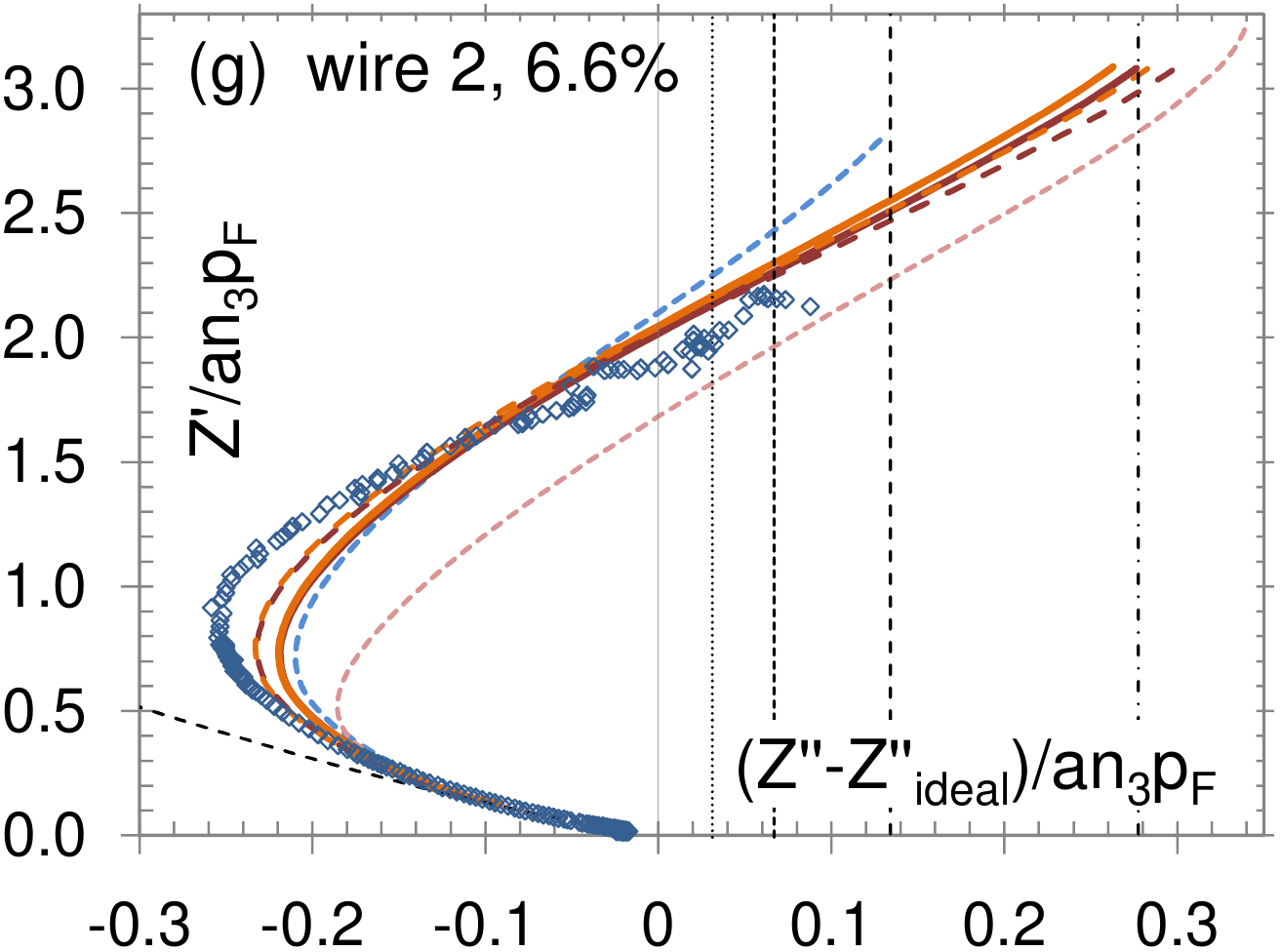}
\includegraphics[width=0.3\linewidth]{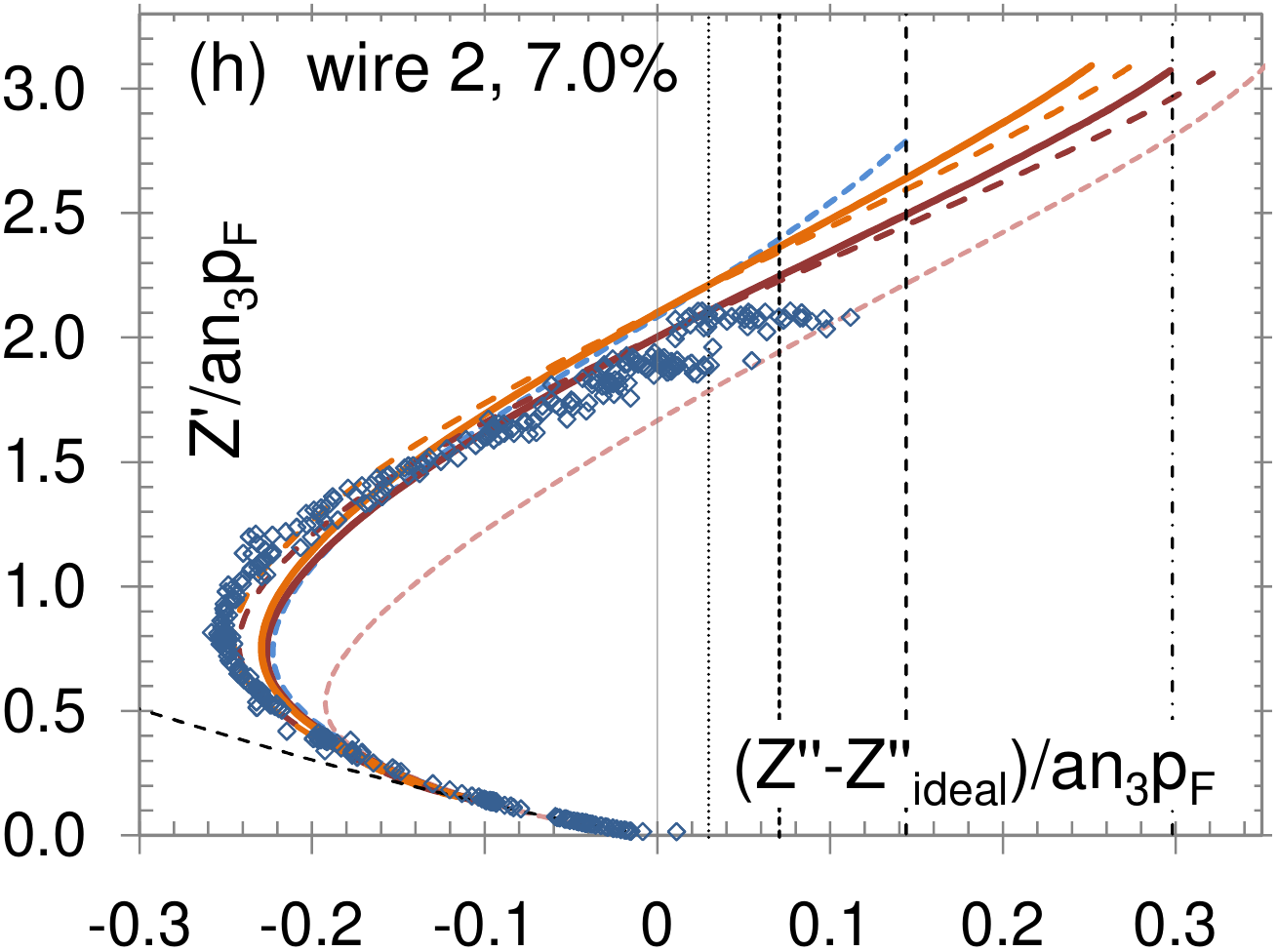}
\includegraphics[width=0.3\linewidth]{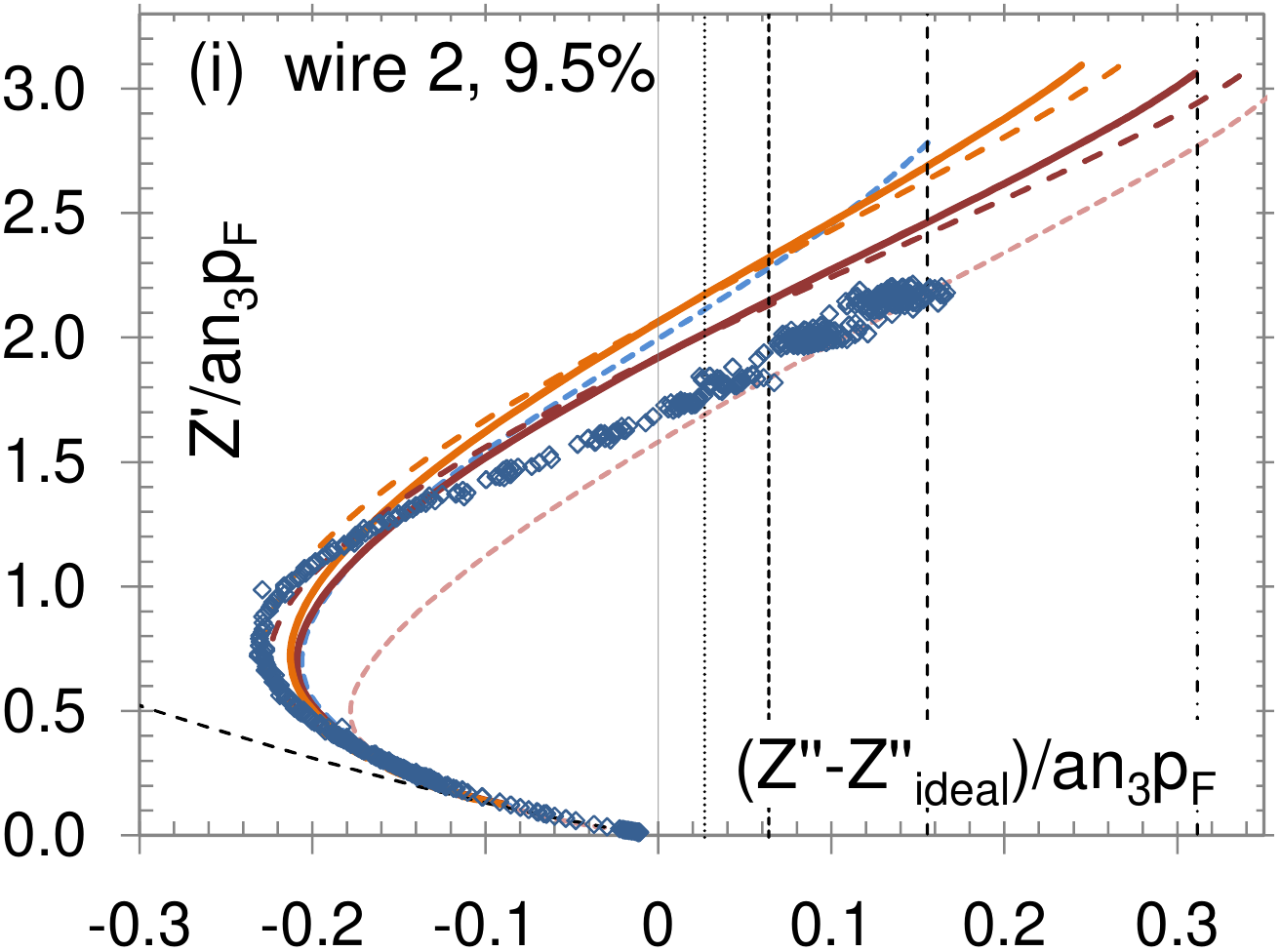}
\caption{Comparison of theory to measurements\cite{mart,Pentti09}. Wire 1 has been measured at six concentrations (1.7\%, 3.6\%, 5.6\%, 6.6\%, 7.0\% and 9.5\%) and wire 2 at the three highest of these. At concentrations 7.0\% and 9.5\% the pressure is 10 atm, while at the others it is saturated vapor pressure. The experimental results are shown by blue data points. The principal theoretical result is the red solid line in each panel. It corresponds to the slab geometry with diffusive walls, $F_0$ from the first row of Table \ref{tab5} and other parameters from the left side of Table \ref{tab4}. Other lines differ from it as follows. The orange lines use $F_0$ from the second row of Table \ref{tab5}. The dashed red and orange lines use the right side of Table \ref{tab4}. The dashed blue line has absorbing container walls.  The dashed pink line has the cylindrical container of radius $b=8a$. The dashed black line is the hydrodynamic result in unlimited fluid. The vertical black lines give the effects 1-4, see text. }\label{fig-expr}
\end{figure}

Our calculations are compared to experiments of Martikainen \emph{et al.}\cite{mart,Pentti09} in Fig.\ \ref{fig-expr}. We model the experimental chamber as the slab container with diffusive boundary condition  on both the wire and the container.  Since there is uncertainty in $m^*$ and $F_0$, we show four combinations of parameters for each concentration. We also show results for the absorbing boundary condition at the container walls and for the cylindrical container of radius $b=8a$ with the diffusive boundary condition. Although the agreement between the experiments and the calculations is not perfect, it is for most cases satisfactory, noting that no fitting parameters have been used.

In the ballistic limit the resonance frequency increases to value higher than in the high temperature limit. This overshoot was analyzed in detail in Ref.\ \onlinecite{VTletter}. There the overshoot was divided into four different contributions: 1) The $^3$He part of the fluid decouples from the ideal fluid flow around the wire. 2) Part of $^4$He moves with the quasiparticles and therefore is also decoupled. [These two contributions correspond to the normal density $\rho_n$ being subtracted from the total density $\rho$ appearing in (\ref{Z-ideal})]. 3) The Landau force due to quasiparticles interactions, caused mainly by $F_0$, adds elasticity to the Fermi liquid and thus increases the resonant frequency. 4) There are corrections caused by the finite size of the container, in particular the effect of quasiparticles reflected from the container back to the wire (\ref{e.foraabd}). The four contributions are shown by vertical lines in Fig.\ \ref{fig-expr}.

We see that for diffuse chamber walls the calculated overshoot is larger than in the experiments. It is possible that the ballistic limit was not quite reached in the experiments. Alternatively, the calculations with absorbing container walls give overshoot of the same magnitude as the experiments. The experimental cell was surrounded by porous sintered silver, which is likely to absorb some of the quasiparticles rather than reflecting them. We note that the slope of the ballistic branches of the calculated curves would better fit to their experimental counterparts if  larger values of $-F_0$ were used. Also, allowing non-zero specular-scattering fraction $S$ would scale down the calculated curves, giving better fit for some of the concentrations, but a systematic improvement is hard to obtain. We point out that for the calculations in cylindrical geometry we have fixed the radius $b=8a$ rather arbitrarily: using $b$ as a fitting parameter would certainly lead to better agreement with experiments, but again it is difficult to find a single value that would fit all the concentrations.

\begin{figure}[!tb]
\centerline{
\includegraphics[width=0.3\linewidth]{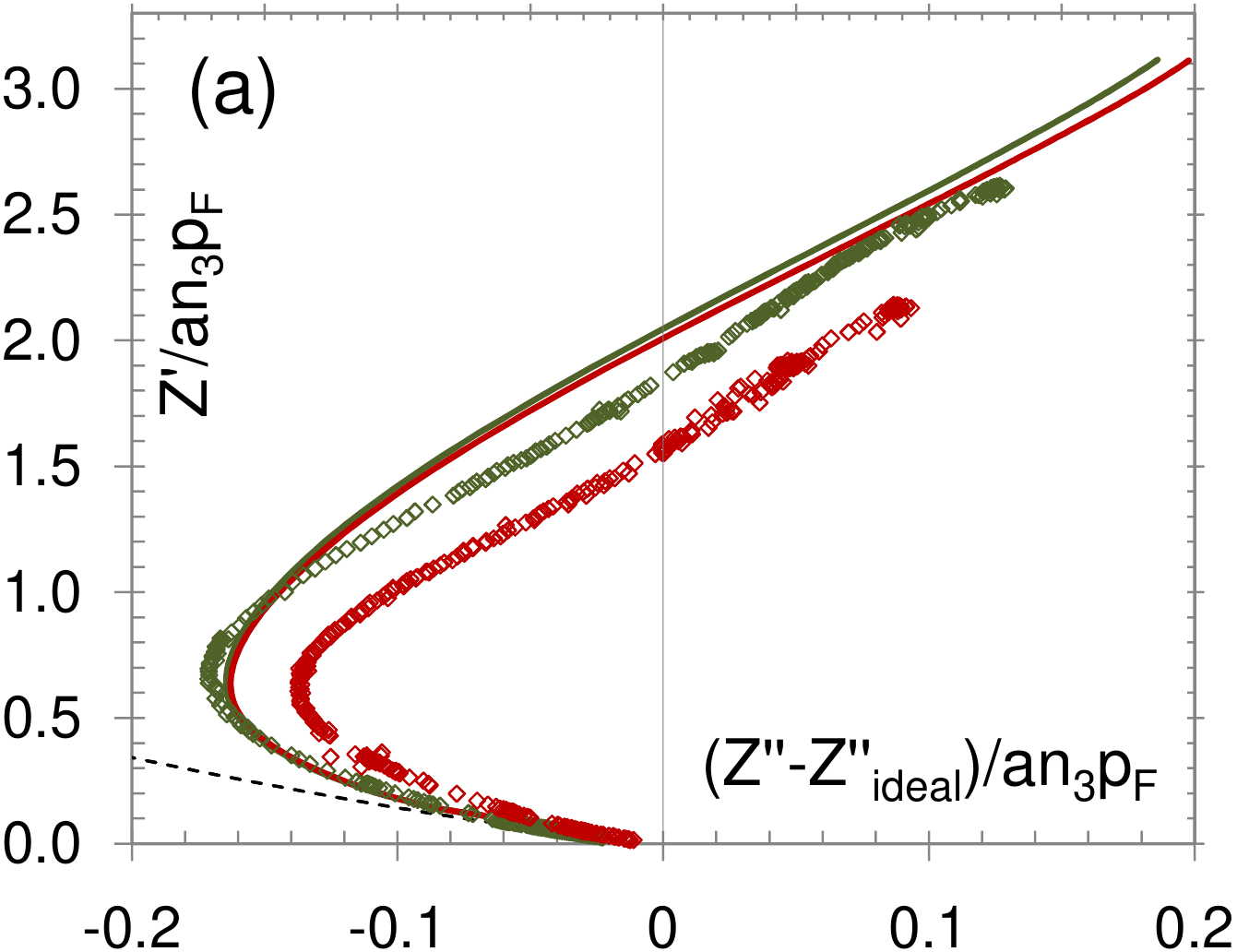}
\includegraphics[width=0.3\linewidth]{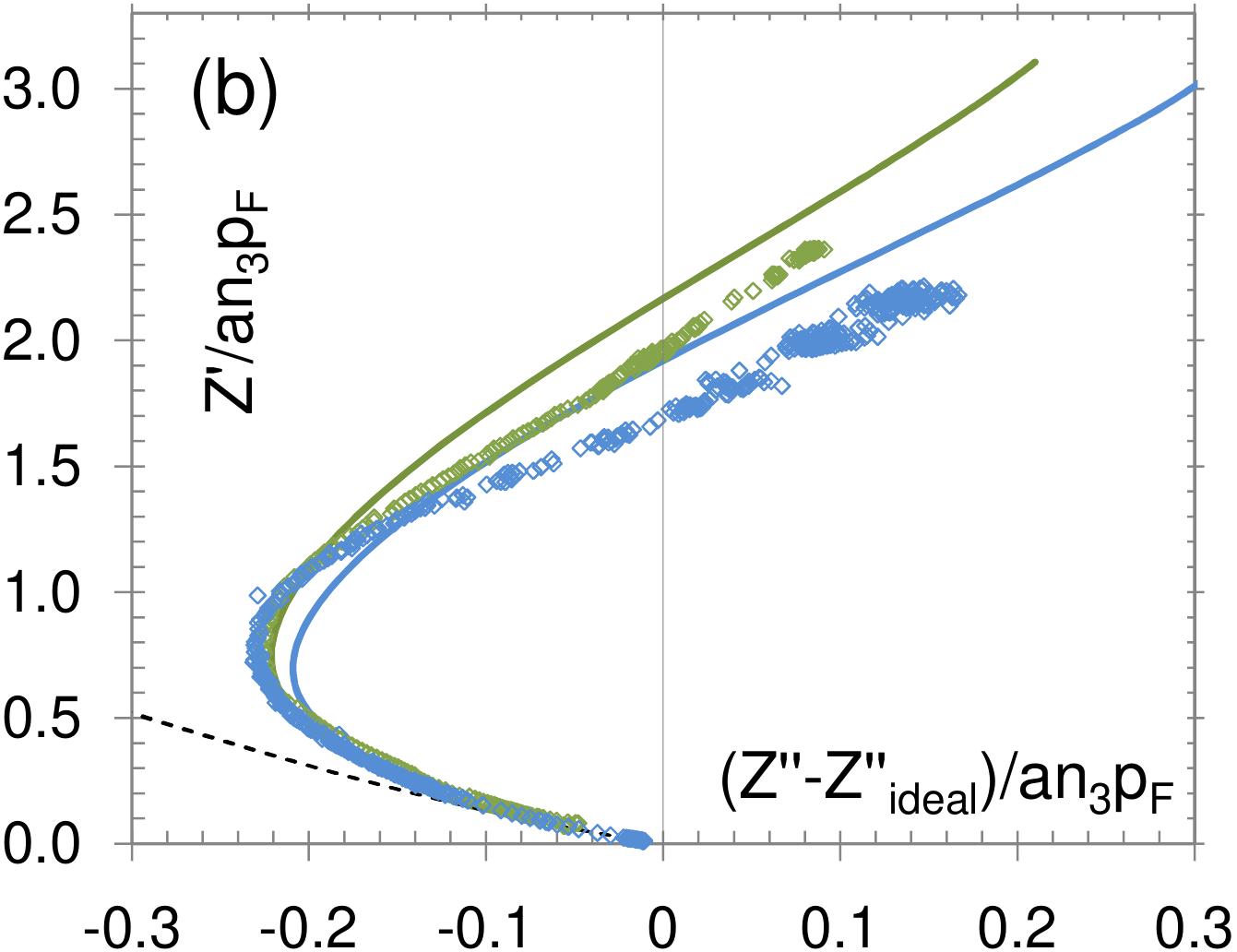}
} \caption{(a) The experimental (circles) and numerical results (solid lines) for wire 1 for two concentrations, 5.6\% at saturated vapor pressure (green) and 7.0\% at $P=10$ atm (red). The hydrodynamic results (dashed lines) are indistinguishable on this scale. We see that the numerical results differ only slightly from each other in the ballistic limit. The reason for the difference between the experimental results is unknown.
(b) The frequency parameters for wire 1, $x_3=1.8\%$ curve (green) and wire 2, $x_3=9.5\%$ curve (blue) are nearly equal, $\Omega \approx0.022$. The numerical curves (solid lines) overlap in the hydrodynamic region, but start to differ around the bend in the curve, due to different values for $F_1$. The slopes of the calculated curves in the ballistic regime differ because of different values for $F_0$.
The theoretical curves are identical to the principal curves in Fig.\ \ref{fig-expr}.}
\label{fig-exp} 
\end{figure}

We see from table \ref{tab4} that for concentrations 5.6\% and 7.0\% for wire 1, the frequency parameters are nearly the same, $\Omega\approx0.016$ (or $\Omega\approx0.017$ depending on the choice of $m_H$). However, the experimental results, converted to the $(Z'',Z')$-curve, differ considerably from each other, as seen in Fig.\ \ref{fig-exp}a. The corresponding numerically calculated curves differ only slightly, due to small difference in $F_0$. There seems to be some problem involved with the two curves at higher pressure for wire 1, and for these the fit to numerical calculations is by far poorest of the nine cases, as seen in Fig. \ref{fig-expr}e and f.

Another interesting coincidence is that for the 1.8\% curve of wire 1 and for the 9.5\% curve for wire 2, the frequency parameters are similar $\Omega\approx0.022$ (or $\Omega\approx0.024$). The experimental curves shown in Fig.\ \ref{fig-exp}b are nearly  identical in the hydrodynamic region, and only start to differ at $\ell>a$. The slopes of the ballistic branches of the curves are different, which is conveniently explained by the different values of $F_0$ for the two concentrations.

There is a systematic difference between the results of wires 1 and 2. For wire 1 the calculated minimum frequencies are smaller than experimentally observed, while for wire 2 the opposite is true. As already discussed, the high pressure curves (7\% and 9.5\%) for wire 1 seem anomalous. 
There also seems to be fine structure in the measured curves that is not present in the calculations. In particular, the nearly straight part from the bend of the curves to the ballistic limit  seems to have small positive curvature in the measured curves. As shown above, the theory  gives second-sound resonances, but at higher frequencies or at larger wall distances than used in the measurements.
In other directions the experimental container is ten times larger and therefore second-sound resonances in these directions should take place at the experimental frequencies. Ideally, such modes are not coupled to the oscillation of the wire, but there might be coupling  if the wire is not perfectly aligned with the slab. This  may be the origin of the observed structures, which are not reproduced by our ideal infinite-cylinder model.

Our calculation can be generalized in  a couple of relatively simple ways. We can tilt the oscillation direction in the slab, we can allow cylindrical containers of more complicated cross section, and we can allow two relaxation times (one for $l=2$ spherical harmonics of $\psi_{\hat{\bm p}}$ and another for higher harmonics). We have tested all these, but they do not seem to give any obvious improvement in the comparison above. Further possible generalizations could be more exact treatment of the collision term, including higher Fermi-liquid parameters like $F_2$, and allowing more general boundary conditions where degree of specularity depends on the angle of an incident quasiparticle. It seems unlikely, though, that these could lead to much better understanding of the experiments.

\section{Conclusion}

We have calculated the response of Fermi-Bose liquid to an oscillating cylinder. This is applied to vibrating wires in $^3$He-$^4$He mixtures. The results differ from ideal-gas results because of Fermi-liquid effects. In particular, the resonance frequency in the ballistic limit exceeds the ideal-fluid value because of $^4$He bound to quasiparticles and the Landau force, i.e.\ the elasticity of the Fermi liquid caused by interactions between the quasiparticles. The results are compared to measurements. For the comparison it is essential to take into account the size and the form of the experimental container. Good agreement is achieved without any fitting parameters.  It seems that to explain the remaining differences would require 3D simulation of experimental volume.

\section*{Acknowledgments}

We thank N. Kopnin, E. Pentti, P. Pietil\"ainen, M. Saarela, J. Tuoriniemi and G. Volovik for useful discussions. We thank the Academy of Finland,  the Finnish Cultural Foundation and the National Graduate School in Materials Physics for financial support.

\end{document}